\documentclass[journal=jpcafh,manuscript=article]{achemso}

\usepackage[version=4]{mhchem} % Formula subscripts using \ce{}
\usepackage{color}
\usepackage[nolist]{acronym}
\graphicspath{{figures/}}
\usepackage{xcolor}


\def\rmdj {d\llap{\raise 1.22ex\hbox
  {\vrule height 0.09ex width 0.315em}\kern 0.04em}}
\def\sldj {d\llap{\raise 1.22ex\hbox
  {\vrule height 0.09ex width 0.265em}}\rlap{\raise 1.22ex\hbox
  {\vrule height 0.09ex width 0.05em}}}
\def\itdj {d\llap{\raise 1.22ex\hbox
  {\vrule height 0.09ex width 0.2em}}\rlap{\raise 1.22ex\hbox
  {\vrule height 0.09ex width 0.06em}}}
\def\bfdj {d\llap{\raise 1.16ex\hbox
  {\vrule height 0.126ex width 0.308em}\kern 0.04em}}
\def\ttdj {\rlap{\kern 0.17em\raise 1.1ex\hbox
  {\vrule height 0.09ex width 0.295em}}d}
\def\scdj {\rlap{\kern 0.04em\raise 0.57ex\hbox
  {\vrule height 0.09ex width 0.20em}}d}
\def\sfdj {d\llap{\raise 1.22ex\hbox
  {\vrule height 0.10ex width 0.3em}\kern 0.02em}}

\def\dj{\ifcase\fam \rmdj \or \or \or
  \or \itdj \or \sldj \or \bfdj \or \ttdj \or \sfdj \or \scdj \else \rmdj \fi}

\def\rmDj {\rlap{\kern 0.05em\raise 0.76ex\hbox
  {\vrule height 0.10ex width 0.28em}}D}
\def\slDj {\rlap{\kern 0.1em\raise 0.76ex\hbox
  {\vrule height 0.1ex width 0.28em}}D}
\def\itDj {\rlap{\kern 0.145em\raise 0.76ex\hbox
  {\vrule height 0.1ex width 0.274em}}D}
\def\bfDj {\rlap{\kern 0.044em\raise 0.72ex\hbox
  {\vrule height 0.126ex width 0.287em}}D}
\def\ttDj {\rlap{\kern 0.02em\raise 0.67ex\hbox
  {\vrule height 0.105ex width 0.20em}}D}
\def\scDj {\rlap{\kern 0.08em\raise 0.73ex\hbox
  {\vrule height 0.12ex width 0.24em}}D}
\def\sfDj {\rlap{\kern 0.02em\raise 0.727ex\hbox
  {\vrule height 0.126ex width 0.26em}}D}

\def\Dj{\ifcase\fam \rmDj \or \or \or
  \or \itDj \or \slDj \or \bfDj \or \ttDj \or \sfDj \or \scDj \else \rmDj \fi}

\newcommand{\cs}{\'{c}}

\newcommand{\s}{\v{s}}

\SectionNumbersOn

\author{Antonio Prlj}
\email{antonio.prlj@irb.hr}
\affiliation[Bristol University]
{Centre for Computational Chemistry, School of Chemistry, University of Bristol, Bristol BS8 1TS, United Kingdom}
\alsoaffiliation[Rudjer Boskovic Institute]
{Division of Physical Chemistry, Ru{\dj}er Bo{\s}kovi{\cs} Institute, 10000 Zagreb, Croatia}
\author{Daniel Hollas}
\email{daniel.hollas@bristol.ac.uk}
\affiliation[Bristol University]
{Centre for Computational Chemistry, School of Chemistry, University of Bristol, Bristol BS8 1TS, United Kingdom}
\author{Basile F. E. Curchod}
\email{basile.curchod@bristol.ac.uk}
\affiliation[Bristol University]
{Centre for Computational Chemistry, School of Chemistry, University of Bristol, Bristol BS8 1TS, United Kingdom}

\title{Deciphering the Influence of Ground-State Distributions on the Calculation of Photolysis Observables}

\begin{document}

\begin{acronym}
     \acro{TSH}{trajectory surface hopping}
     \acro{VOCs}{volatile organic compounds}
     \acro{NEA}{nuclear ensemble approach}
     \acro{MHP}{methylhydroperoxide}
     \acro{QT}{quantum thermostat}
     \acro{ICs}{initial conditions}
\end{acronym}

\begin{abstract}

Nonadiabatic molecular dynamics offers a powerful tool for studying the photochemistry of molecular systems.
Key to any nonadiabatic molecular dynamics simulation is the definition of its \textit{initial conditions}, ideally representing the initial molecular quantum state of the system of interest. 
In this work, we provide a detailed analysis of how initial conditions may influence the calculation of experimental observables by focusing on the photochemistry of methylhydroperoxide, the simplest and  most abundant organic peroxide in our atmosphere. We investigate the outcome of trajectory surface hopping simulations for distinct sets of initial conditions sampled from different approximate quantum distributions, namely harmonic Wigner functions 
and \textit{ab initio} molecular dynamics using a quantum thermostat.
Calculating photoabsorption cross-sections, quantum yields, and translational kinetic energy maps from the results of these simulations reveals the significant effect of the initial conditions, in particular when low-frequency ($\sim$ a few hundred cm$^{-1}$) normal modes are connected to the photophysics of the molecule. Overall, our results indicate that sampling initial conditions from \textit{ab initio} molecular dynamics using a quantum thermostat is preferable for flexible molecules with photoactive low-frequency modes.
From a photochemical perspective, our nonadiabatic dynamics simulations offer an explanation for a low-energy tail observed at high excitation energy in the translational kinetic energy map of methylhydroperoxide.

\end{abstract}

\begin{tocentry} 
\includegraphics[width=1.0\textwidth]{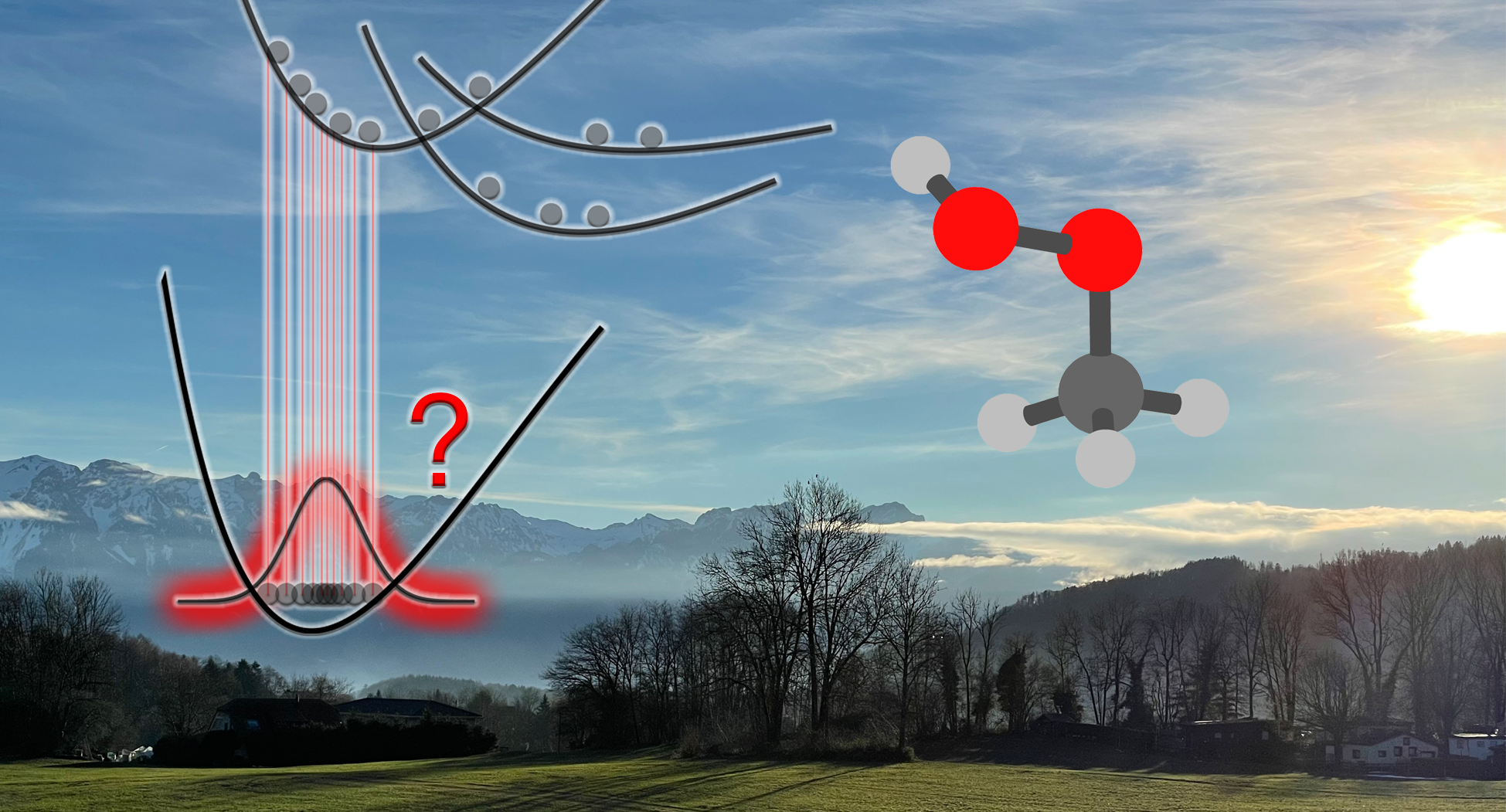}
\end{tocentry}

%%%%%%%%%%%%%%%%%%%%%%%%%%%%%%%%%%%%%%%%%%%%%%%%%%%%%%%%%%%%%%%%%%%%%
%% Start the main part of the manuscript here.
%%%%%%%%%%%%%%%%%%%%%%%%%%%%%%%%%%%%%%%%%%%%%%%%%%%%%%%%%%%%%%%%%%%%%
\newpage
\section{Introduction}
Nonadiabatic molecular dynamics has become a widely used tool to explore molecular photochemistry and predict relevant photochemical observables. Its applications range, for example, from time-resolved photoelectron spectroscopy,\cite{glover2018excited,pathak2020tracking} electron diffraction,\cite{champenois2021conformer,yang2020simultaneous} or ultrafast X-ray scattering\cite{kirrander2016ultrafast,li2017ultrafast} to time-independent observables such as translational energy distributions\cite{fu2011three} and quantum yields of photochemical reactions.\cite{granucci2007excited,thompson2018first} The \textit{in silico} prediction of experimental observables is particularly useful for molecular systems where experimental measurements are difficult to interpret or sometimes even challenging to conduct. 
A typical example of the latter is given by transient \ac{VOCs}. These molecules are of key importance for atmospheric chemistry\cite{ravishankara2015,atkinson2003atmospheric,vereecken2015theoretical}, but they are notoriously difficult to study experimentally due to their reactivity and short lifetime. As some \ac{VOCs} can interact with light and undergo photolysis, \textit{ab initio} photochemical tools can be readily employed to estimate observables that are important in atmospheric modeling.\cite{prlj2020theoretical}
In particular, the rate coefficient $J$ for a first-order photolytic process can be evaluated as $J = \int_{\lambda_{min}}^{\lambda_{max}} \sigma (\lambda) \phi (\lambda) F(\lambda) d\lambda$, where $\sigma (\lambda)$ is the photoabsorption cross-section of the molecule, $\phi (\lambda)$ the photolysis wavelength-dependent quantum yield, $F(\lambda)$ the photon flux of the light source (actinic flux when the source is sunlight), and $\lambda$ the wavelength. Both $\sigma (\lambda)$ and $\phi (\lambda)$ are time-independent observables that can be predicted with state-of-the-art computational methods. Our group has recently proposed a protocol\cite{prlj2020theoretical} to determine $\sigma (\lambda)$ using the \ac{NEA}\cite{crespo2012spectrum} and $\phi (\lambda)$ by resorting to \ac{TSH} dynamics.\cite{tully1990molecular} This protocol has been applied to \ac{VOCs} like \textit{tert}-butyl hydroperoxide,\cite{prlj2020theoretical} 2-hydroperoxypropanal,\cite{marsili2022theoretical} or pyruvic acid.\cite{hutton2022photodynamics} TSH simulations and the NEA were also used to successfully unravel the photochemistry of other atmospheric molecules (for examples, see Refs.~\citenum{doi:10.1021/acs.jpca.8b12482,mcgillen2017criegee,D3CP00207A,mccoy2021simple,doi:10.1021/acs.jpca.2c08025,frances2020photodissociation,carmona2021photochemistry}).

The \ac{NEA} and \ac{TSH} are among the most popular computational methods used to study the photophysics and photochemistry of medium-size molecular systems.\cite{crespo2018recent} Both strategies rely on the determination of a ground-state nuclear density to sample a set of discrete phase-space \ac{ICs}, \textit{i.e.}, nuclear coordinates and momenta. By using the geometries of hundreds of \ac{ICs}, the \ac{NEA} proposes to calculate their electronic transitions and broaden them with appropriate shape functions to obtain a convoluted photoabsorption cross-section -- $\sigma (\lambda)$ -- that accounts for non-Condon effects. \ac{TSH} trajectories are commonly initiated from the same pool of \ac{ICs} used for the \ac{NEA}. \ac{TSH} is a mixed quantum-classical approach in which a swarm of classical trajectories evolves in multiple electronic states with the possibility of inter-state hopping.\cite{tully1990molecular} It can be employed for systems that are excited instantaneously by an ultrashort laser pulse as well as by continuum-wave fields such as solar irradiation.\cite{barbatti2020simulation}  As trajectories may have various fates and yield various photoproducts, the different resulting quantum yields, $\phi$, are typically evaluated by counting the trajectories giving a certain product and dividing their total number ($N_{product}$) by the total number of trajectories in a swarm ($N_{tot}$). If wavelength-dependent quantum yields, $\phi (\lambda)$, are needed, the \ac{ICs} for \ac{TSH} dynamics can be selected from narrow excitation-energy windows in $\sigma (\lambda)$, centered around different $\lambda$ values.\cite{prlj2020theoretical}

What are then the strategies available to map a ground-state nuclear probability density distribution into \ac{ICs}? Earlier works commonly used Boltzmann (thermal) sampling by running a long Born-Oppenheimer ground-state dynamics and taking a large number of snapshots as \ac{ICs} for the \ac{NEA} (geometries) and the \ac{TSH} dynamics (geometries and momenta).
However, as stressed by Barbatti and Sen,\cite{barbatti2016effects} Boltzmann sampling does not fully recover the zero-point energy, resulting in \ac{NEA} absorption bands that are typically too narrow when compared to experimental data. The outcome of \ac{TSH} dynamics can also be affected by Boltzmann sampling, leading, for instance, to an extension of the timescales of nonradiative decay and a change in the distribution of reaction pathways.\cite{barbatti2016effects} A Wigner distribution is a more rigorous way to map quantum nuclear densities on quasi phase-space quantities, recovering the quantum delocalization of nuclei and zero-point vibrational effects naturally. Sampling \ac{ICs} from a Wigner distribution is a common strategy used in many recent works related to excited-state dynamics, but its drawbacks have also been scrutinized.\cite{persico2014overview,mccoy2014role,suchan2018importance, mai2018novel} When dealing with realistic multidimensional molecular systems, the Wigner distribution is commonly implemented within the harmonic approximation for uncoupled normal modes, which restricts its reliability to molecular systems with limited anharmonicity. Furthermore, linear normal modes poorly represent torsional degrees of freedom, typically resulting in light atoms being artificially displaced.\cite{mccoy2014role,mai2018novel,persico2014overview} Viable \textit{ad hoc} corrections resort to filtering out the `problematic' low-frequency modes from the Wigner distribution.\cite{svoboda2011simulations,favero2013dynamics}

As an alternative to Wigner sampling, Suchan et al.\cite{suchan2018importance} advocated the use of a \ac{QT} in ground-state dynamics to sample \ac{ICs}. \ac{QT}\cite{ceriotti2009nuclear,ceriotti2010colored,Finocchi2022} is based on a generalized Langevin equation (GLE) thermostat that keeps the normal modes of a molecular system at different frequency-dependent temperatures -- as such, \ac{QT} provides phase-space distributions corresponding to quantum harmonic oscillators. \ac{QT} can properly treat both high and low-frequency modes and performs well even for (moderately) anharmonic systems.\cite{suchan2018importance,Vuilleumier2013} We note that a different implementation of a similar idea was proposed by Dammak et al.\cite{Dammak2009} and termed quantum thermal bath (QTB). QTB was applied to studying vibrational spectra\cite{Bonella2021} or the structure of liquid water.\cite{Mauger2021,Ple2023} A recent extension of QTB was devised to tackle the zero-point energy leakage issue.\cite{Brieuc2016,Mangaud2019} Our group has recently compared the impact of \ac{QT} and Wigner sampling on the prediction of photoabsorption cross-sections of several exemplary \ac{VOCs} within the \ac{NEA}.\cite{prlj2021calculating} \ac{QT} was found superior whenever low-frequency anharmonic modes play a role in the photochemistry/photophysics of the molecule.

One can ask a reasonable question at this stage: what is the influence of the different strategies to sample \ac{ICs} on the observables predicted by \textit{ab initio} simulations? In this work, we propose to investigate the impact that Wigner and \ac{QT} sampling strategies may have on the prediction of a series of experimental observables for \ac{MHP} --- \ce{CH3OOH}. \ac{MHP} is a VOC relevant to atmospheric chemistry\cite{wang2023organic} that, despite its simple structure, poses numerous challenges to computational photochemistry. 
The observables of interest in this work are photoabsorption cross-sections ($\sigma(\lambda)$), wavelength-dependent quantum yields ($\phi (\lambda)$), and translational kinetic energy distributions, predicted from the \ac{NEA} and \ac{TSH} simulations based on XMS-CASPT2 electronic structure (see Computational Details for further information).

\section{Computational Details}
\subsection{Electronic-structure methods}
\begin{figure}[!ht]
\includegraphics[width=0.5\textwidth]{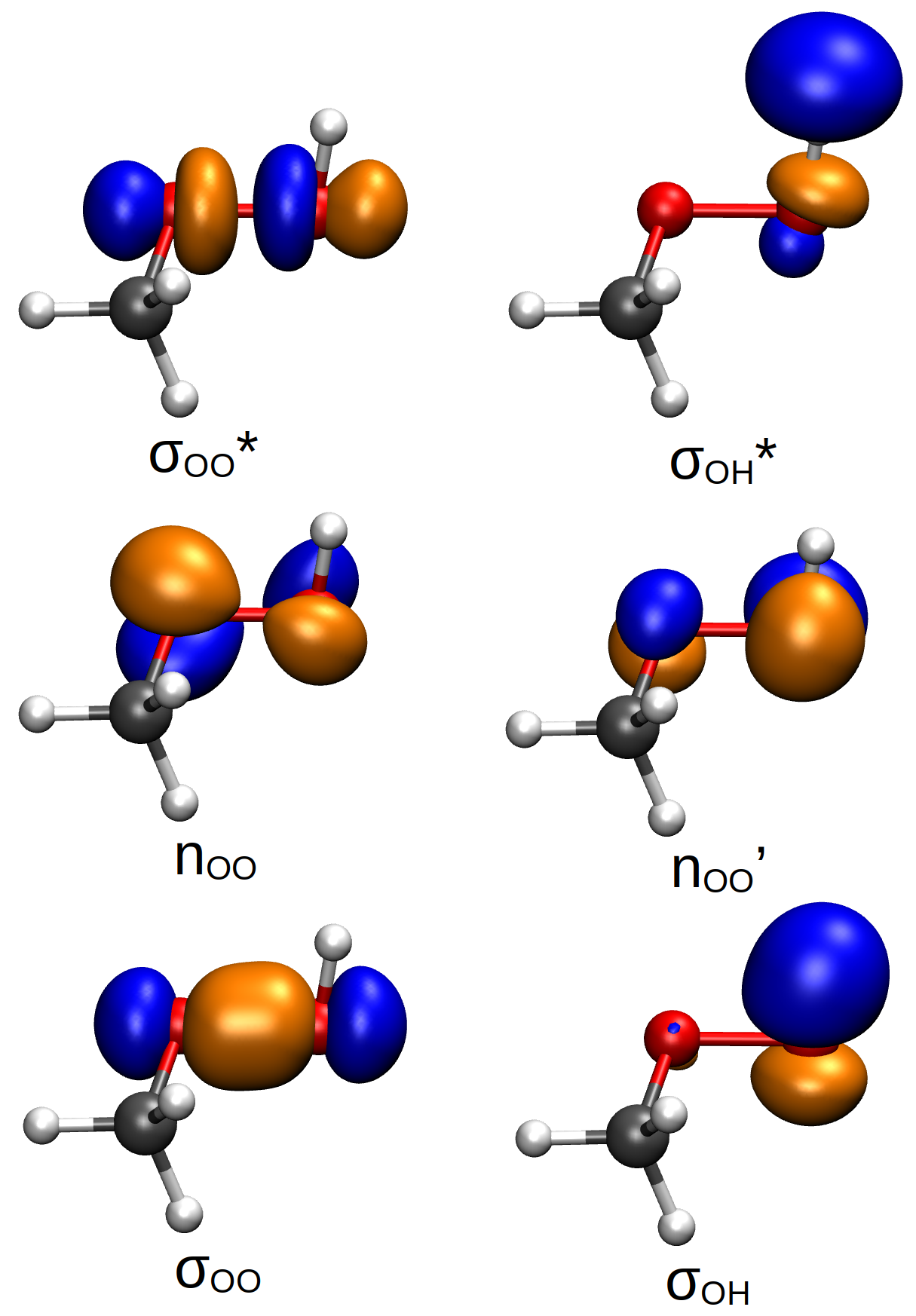}
  \caption{Active space orbitals employed in the XMS(4)-CASPT2(8/6)/def2-SVPD calculations, given here for the ground-state optimized geometry obtained with MP2/aug-cc-pVDZ. Isovalue was set to 0.1.}
  \label{fgr:Fig5}
\end{figure}

The ground and three lowest excited electronic singlet states of MHP were calculated with extended multi-state complete active space second-order perturbation theory (XMS-CASPT2)~\cite{shiozaki2011communication} using BAGEL 1.2.0 package.~\cite{shiozaki2018bagel}  The choice of the multireference method XMS-CASPT2 is dictated by the fact that the photodissociation dynamics (i.e., bond breaking) of a molecule bearing a hydroperoxy group cannot be properly described by a single-reference method such as LR-TDDFT or ADC(2).\cite{prlj2020theoretical} We employed XMS(4)-CASPT2(8/6) along with a def2-SVPD basis set,~\cite{rappoport2010property} where the active space was composed of six orbitals, two nonbonding $n$ orbitals localized on \ce{O} atoms and two pairs of bonding and antibonding $\sigma / \sigma^\ast$ orbitals describing \ce{O-O} and \ce{O-H} bonds (see Fig.~\ref{fgr:Fig5}). 
In contrast to earlier work on tert-butylhydroperoxide~\cite{prlj2020theoretical} (see Fig.~S3 in Ref.~\citenum{prlj2020theoretical}), we excluded the $\sigma / \sigma^\ast$ orbitals of the \ce{C-O} bond as they proved to have no impact on low-lying electronic states and ensuing nonadiabatic molecular dynamics (for both tert-butylhydroperoxide and MHP). 
XMS-CASPT2 was employed within the single state - single reference (SS-SR) contraction scheme.~\cite{vlaisavljevich2016nuclear} A real vertical shift was set to 0.5 a.u. to avoid problems with intruder states. Similar values for the vertical shift were used in earlier XMS-CASPT2-based TSH simulations.\cite{polyak2019ultrafast,prlj2020theoretical} We found that this shift value increases the numerical stability of the TSH dynamics of MHP, even though it slightly deteriorates excitation energies and oscillator strengths (see Table~S3). Frozen core and density-fitting approximations (using the def2-TZVPP-jkfit basis set from the BAGEL library) were applied. A detailed benchmark of the electronic energies and oscillator strengths with other electronic structure methods, including the high-level CC3 reference,\cite{loos2018mountaineering,folkestad2020t} is given in the SI. Orbitals and molecular representations were visualized with the VMD package, version 1.9.3.\cite{humphrey1996vmd}

\subsection{Ground-state sampling and photoabsorption cross-sections}
A harmonic Wigner distribution was calculated with the SHARC 2.1 package,\cite{richter2011sharc,mai2018nonadiabatic} using all the harmonic normal mode frequencies obtained for the ground-state minimum-energy structure. To investigate the effect of low-frequency modes, it is possible to omit certain vibrational modes from the harmonic Wigner sampling. In the specific case of MHP, we performed a Wigner sampling without the lowest-frequency normal mode, corresponding to the \ce{C-O-O-H} torsion. We refer to this distribution as Wigner*.
Geometry optimizations and normal-mode calculations were performed with Turbomole 7.4.1.~\cite{furche2014turbomole} at the MP2/aug-cc-pVDZ level of theory (see MP2 benchmark in Ref.~\citenum{watts2006ground}, as well as SI of Ref.~\citenum{prlj2020theoretical}). MP2/aug-cc-pVDZ provides geometries that are very similar to those obtained with XMS-CASPT2/def2-SVPD,\cite{prlj2020theoretical} while it can be used for a long ground-state dynamics simulations needed for QT sampling. 

QT sampling was performed with the ABIN code,\cite{abin} coupled to Turbomole for the electronic structure. 
The GLE thermostat parameters in the form of drift matrix $\mathbf{A}$ and diffusion matrix $\mathbf{C}$ were taken from the GLE4MD web page,\cite{GLE4MDwebsite} using a target temperature $T$ = 298~K, number of additional degrees of freedom $N_s=6$, $\hbar\omega_{\mathrm{max}}/kT=20$, and using the strong coupling regime to prevent issues with zero point energy leakage.\cite{ceriotti2010colored} $\omega_{\mathrm{max}}$ corresponds to the maximum normal mode frequency for which the GLE parameters were optimized for a given temperature $T$. For $T$ = 298 K, the maximum frequency evaluates to 4114~cm$^{-1}$, which is well above the largest frequency in MHP (3756~cm$^{-1}$ at the MP2/aug-cc-pVDZ level of theory).
A time step of $\sim$0.5~fs was used in molecular dynamics, and the equilibration time was determined by monitoring the convergence of the average kinetic energy temperature. To benchmark the QT distributions in the coordinate space, we also performed path integral simulations combined with the GLE thermostat within the PI+GLE method.\cite{ceriotti2011} This strategy converges to the exact quantum results faster than the canonical PIMD. We used four path-integral beads, while the PI+GLE parameters were again taken from the GLE4MD web page, using $T$ = 298~K and parameters: $N_s=6$, $\hbar\omega_{\mathrm{max}}/kT=50$.

In total, 4000~\ac{ICs} were selected for each type of sampling (Wigner, Wigner*, and QT). Electronic excitation energies for the three lowest singlet excited states and their oscillator strengths were calculated with XMS(4)-CASPT2(8/6)/def2-SVPD. A small fraction of \ac{ICs} had to be discarded due to issues with electronic-structure convergence (see SI for details). Absolute photoabsorption cross-sections were calculated within the NEA as implemented in the Newton-X 2.0 package,\cite{barbatti2014newton,barbatti2022newton} using a phenomenological Lorentzian broadening of 0.05~eV. 

\subsection{Excited-state molecular dynamics}

TSH~\cite{tully1990molecular} simulations were performed with SHARC 2.1,\cite{richter2011sharc, mai2018nonadiabatic} interfaced with BAGEL for the electronic-structure calculations. The TSH dynamics involved four singlet electronic states and TSH trajectories were typically 25~fs long -- the timescale was extended up to 100~fs for a small number of trajectories where the photolysis outcome was unclear within the first 25~fs. The time step for the nuclear dynamics was 0.5~fs, with 25 substeps for the propagation of the electronic quantities. The decoherence correction devised by \citeauthor{granucci2007critical}\cite{granucci2007critical} was used to correct the TSH electronic populations. Nonadiabatic couplings were calculated with the wavefunction overlap scheme. After a successful hop, the kinetic energy was adjusted by rescaling the nuclear velocity vector isotropically. 
For each type of sampling (Wigner, Wigner*, QT), the TSH dynamics was initiated from a subset of the 4000 initial conditions used for the NEA (see Table~S1 for details about the numbers of TSH trajectories). We defined three narrow energy windows within each photoabsorption cross-section, centered around 5.00~eV (248~nm), 5.71~eV (217~nm), and 6.42~eV (193~nm). Each window had a total width of 0.3~eV. \ac{ICs} were selected within a window if their transition energies fell within the energy range of the window. The $f$-biased selection scheme employed for some TSH simulations was applied by modifying the \texttt{excite.py} script in SHARC 2.1 (see details in Sec.~\ref{resultdiscussion}). We implemented an $f$-biased selection with excitation probabilities proportional solely to the oscillator strengths. To calculate translational kinetic energy maps, the nuclear velocities of \ce{OH} and \ce{CH3O} fragments were collected after 25 fs of dynamics. No special treatment against zero-point energy (ZPE) leakage was applied given the very short timescale of the TSH simulations reported in this work (for a detailed discussion of the effect of ZPE in nonadiabatic dynamics, the interested reader is referred to Ref.~\citenum{doi:10.1021/acs.jctc.3c00024}). The reader is referred to the SI for a comment about the discarded trajectories and their potential impact on the calculated quantum yields and kinetic energy maps.

\section{Results and Discussion}
\label{resultdiscussion}
\subsection{Approximate ground-state nuclear density of methylhydroperoxide}
\label{secgeom}
\begin{figure}[!ht]
\centering
\includegraphics[width=1.0\textwidth]{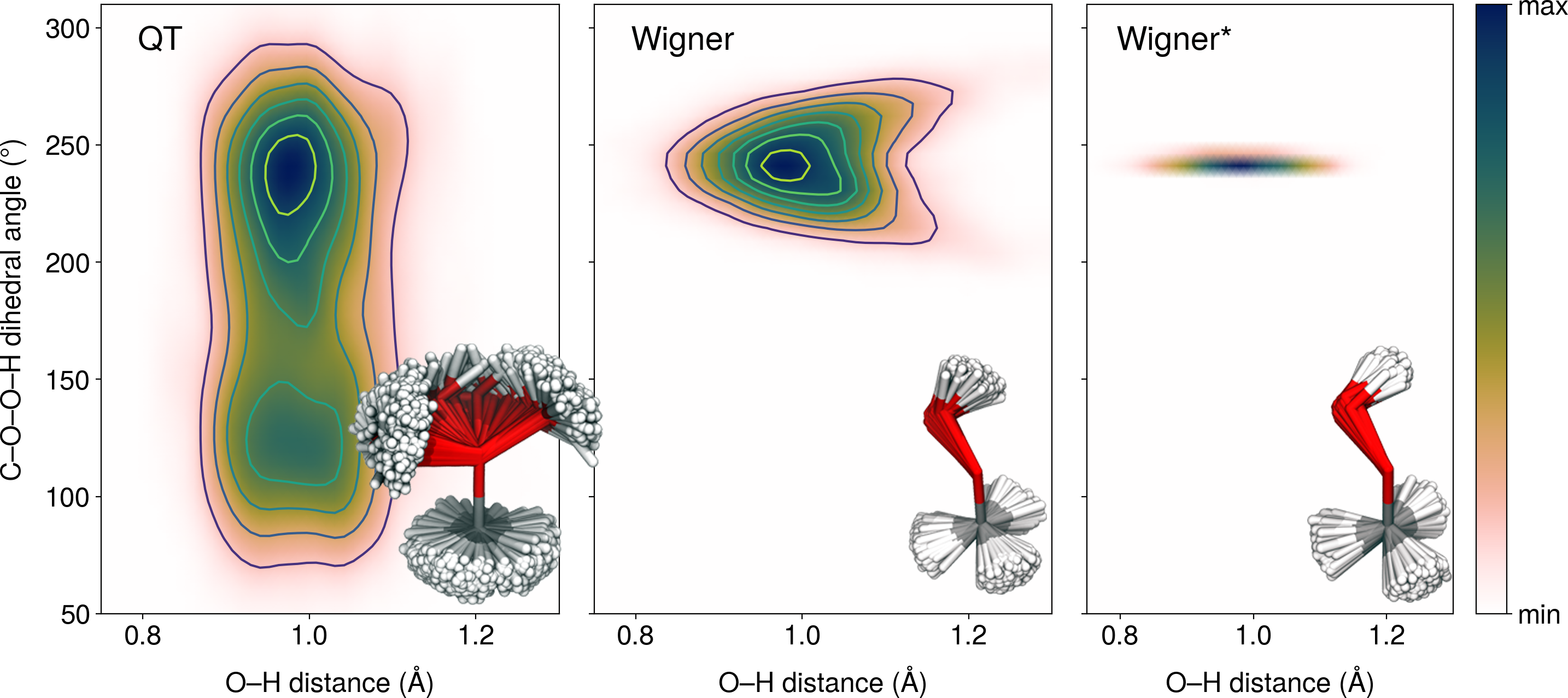}
\caption{Distribution of the \ce{O-H} distance and \ce{C-O-O-H} dihedral angle of MHP for 4000 geometries sampled from: (left panel) an \textit{ab initio} molecular dynamics (MP2/aug-cc-pVDZ) using a quantum thermostat, (middle panel) a Wigner distribution for uncoupled harmonic oscillators obtained from the MP2/aug-cc-pVDZ equilibrium geometry of MHP and corresponding harmonic frequencies, and (right panel) the same Wigner distribution but with the low-frequency \ce{C-O-O-H} torsion removed from the sampling. The color maps were created by the Gaussian kernel density estimation using a 50 by 50 grid of points. The kernel bandwidth was estimated using Scott's rule.\cite{Scott1992} The insets show the 4000 geometries sampled from each distribution, aligned with respect to the central \ce{C-O} bond.}
\label{fig:ground-state-distributions}
\end{figure}

\ac{MHP} is the simplest and most abundant organic peroxide in the atmosphere, with implications on atmospheric radical and oxidative balance.\cite{wang2023organic} From a theoretical perspective, \ac{MHP} exhibits an interesting low-frequency normal mode at 201 cm$^-1$ (MP2/aug-cc-pVDZ) that critically affects the sampling of \ac{ICs}.\cite{prlj2021calculating} More specifically, the \ce{C-O-O-H} torsional mode of MHP is poorly sampled when using a distribution built from linear normal modes -- like the Wigner distributions constructed from the equilibrium geometry and vibrational modes obtained from quantum-chemical calculations -- leading to an artificially broad distribution of \ce{O-H} bond lengths. A significant number of MHP geometries sampled from a harmonic Wigner distribution ('Wigner') exhibit \ce{O-H} bond lengths larger than 1.2~\AA~(middle panel, Fig.~\ref{fig:ground-state-distributions}). 
This problem stems from the fact that atoms involved in low-frequency torsions are not \textit{per se} rotated but moved along normal-mode vectors, which causes unphysical displacements of light H atoms. Torsions are inherently curvilinear and notoriously poorly represented by rectilinear normal modes with Cartesian displacements.\cite{prlj2021calculating,mccoy2014role,mai2018novel}
The correlation between \ce{C-O-O-H} torsion and \ce{O-H} bond length is clearly visible in the middle panel of Fig.~\ref{fig:ground-state-distributions}, where torsion along the \ce{C-O-O-H} mode around the equilibrium geometry is connected with an elongation of the \ce{O-H} bond length. We note that sampling the same harmonic Wigner distribution at 300\,K further enhances the artifact observed here at 0\,K (see Fig.\,S1 in the SI). Removing the \ce{C-O-O-H} torsion from the Wigner sampling ('Wigner*') immediately fixes the issue with the \ce{O-H} bond length distribution (right panel of Fig.~\ref{fig:ground-state-distributions} -- see also Fig. S1 in the SI), but we will see later that this strategy is not without danger if the removed mode is of importance for the photochemistry/photophysics studied. 

Sampling geometries from an \textit{ab initio} molecular dynamics with QT leads to a proper distribution of the \ce{C-O-O-H} dihedral angle and \ce{O-H} bond length as these structural parameters are now coupled in the \textit{ab initio} molecular dynamics (left panel of Fig.~\ref{fig:ground-state-distributions}). More importantly, QT reveals the much broader distribution of \ce{C-O-O-H} dihedral angle, with a second energy minimum emerging at $\sim$120$^{\circ}$. We note that this minimum in the ground electronic state of MHP is isoenergetic with that at $\sim$240$^{\circ}$ and a much longer \textit{ab initio} molecular dynamics with QT would be required to reach fully-converged distributions. Structural differences between sampled \ac{ICs} are obvious from the insets in Fig.~\ref{fig:ground-state-distributions} -- \ac{QT} samples almost free rotations around \ce{C-O} and \ce{O-O} bonds, while as mentioned above these rotations are restricted in the Wigner sampling and even frozen in Wigner*. In principle, both minima of the ground-state potential energy surface could be sampled by calculating separate Wigner distributions. This extended sampling is, however, not necessary for the photochemical observables of interest here as the two minima correspond to chemically identical molecules from a symmetry perspective -- with the minimum-energy geometries being mirror images of each other.

To verify the accuracy of QT, we compared its distribution for the \ce{C-O-O-H} torsion angle and \ce{O-H} bond length with fully converged path-integral results obtained with the PI+GLE approach (see Fig.~S1 in the SI). The QT and PI+GLE are in very good agreement, even for the highly anharmonic \ce{C-O-O-H} mode, validating the distributions obtained with QT. 
We note that we also compared the distributions of nuclear momenta between Wigner, Wigner*, and QT (Fig.~S2 in the SI), showing overall a good agreement between the methods. 

The incorrect description of the \ce{O-H} bond length in the Wigner sampling will directly affect the calculated photoabsorption cross-section, $\sigma (\lambda)$, for \ac{MHP}. The sensitivity of $\sigma (\lambda)$ on the accuracy of bond-length distributions lies in the fact that the low-lying singlet excited states of \ac{MHP} exhibit an antibonding $n\sigma^\ast$ character (akin to other alkyl-peroxides\cite{prlj2020theoretical}). The excitation energy of an electronic state exhibiting a $n\sigma^\ast$ character is generally highly sensitive to the length of the chemical bond(s) where the antibonding $\sigma^\ast$ orbital is localized.
At the optimized ground-state geometry of \ac{MHP}, the first excited electronic state (S$_1$) has a $n'\sigma^\ast$(\ce{O-O}) character (see Fig.~\ref{fgr:Fig5} for a depiction of the molecular orbitals). It is followed, approximately 1 eV higher in energy, by two electronic states having a $n\sigma^\ast$(\ce{O-O}) and $n'\sigma^\ast$(\ce{O-H}) (see Table S3 in the SI). Hence, using a proper approximate ground-state nuclear density distribution is critical to ensure an accurate description of the $n'\sigma^\ast$(\ce{O-H}) transition for the calculated photoabsorption cross-section $\sigma (\lambda)$, and also potentially for other observables as we will see below. 

\subsection{Photoabsorption cross-section of methylhydroperoxide}

\begin{figure}[!ht]
\centering
\includegraphics[width=0.9\textwidth]{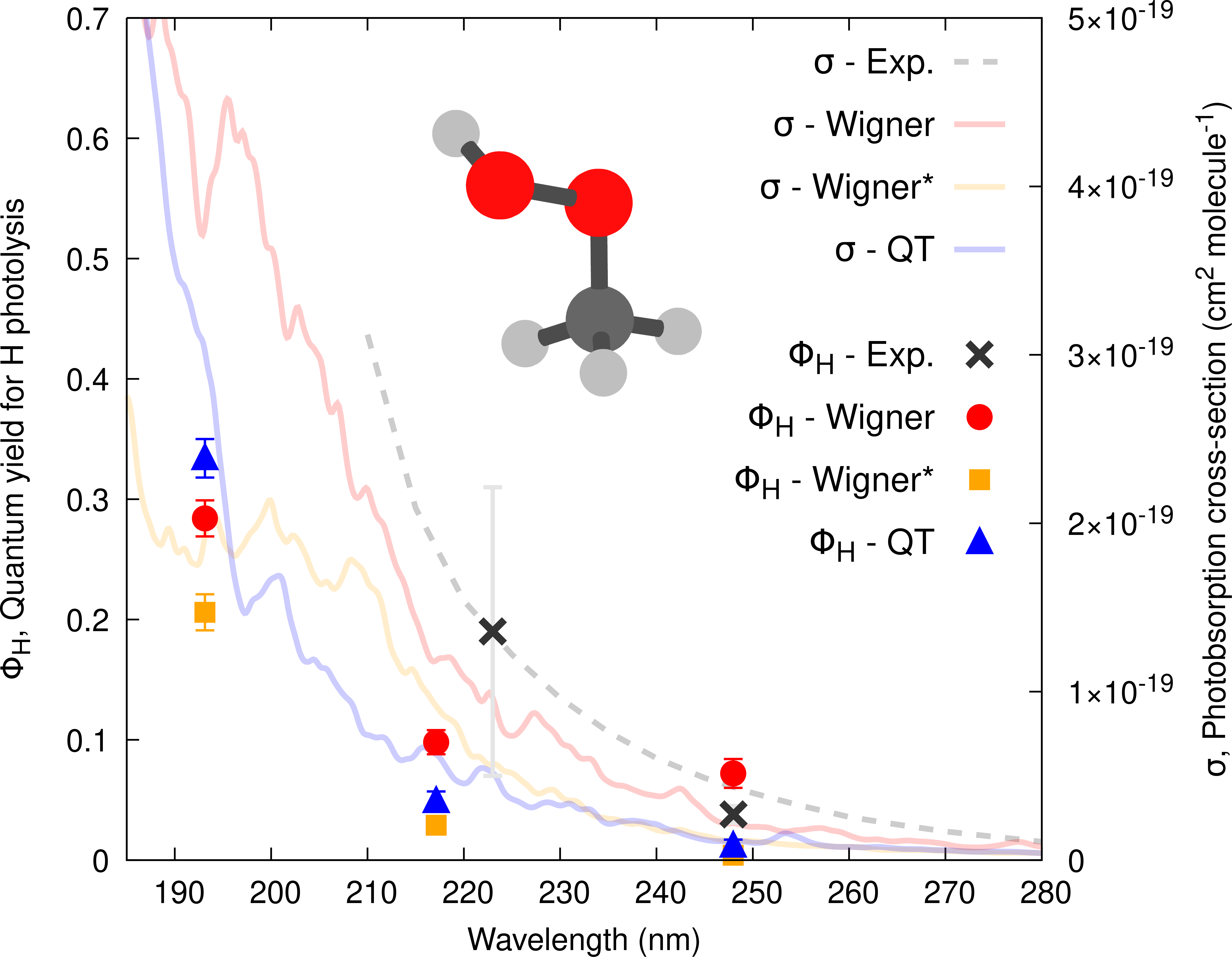}
\caption{Calculated photoabsorption cross-sections ($\sigma$, curves) and wavelength-dependent quantum yields ($\phi_H$, symbols) for the H-atom photolysis from \ac{MHP}. $\sigma$ and $\phi_H$ were obtained from the \ac{NEA} and \ac{TSH} dynamics, respectively, based on an uniform selection of \ac{ICs} generated from a Wigner distribution (red curve and circles), a Wigner distribution with the lowest-frequency mode removed (Wigner$^*$, orange curve and squares), and a \ac{QT}-based \textit{ab initio} molecular dynamics (blue curve and triangles). The dashed curve and black crosses correspond to experimental data (see main text).
Error bars represent the standard deviation for the calculated quantum yields or the reported error bars for the experimental values (obtained from Refs.~\citenum{vaghjiani1990photodissociation,blitz2005wavelength}) -- we note that the quantum yield value at 223 nm is deduced from the quantum yield of the \ce{CH3O} radical (see main text for discussion).  The insets show the molecular structure of MHP.}
\label{fig1}
\end{figure}

Our investigation of the role of IC sampling for photochemical observables begins with the photoabsorption cross-section of \ac{MHP}, $\sigma (\lambda)$. We focus more specifically on the low-energy tail of this quantity as this spectral region plays an important role in the context of atmospheric chemistry due to its overlap with the solar actinic flux. Fig.~\ref{fig1} (right axis) compares the predicted $\sigma (\lambda)$ (colored curves) using the three different sampling procedures for the \ac{NEA} -- Wigner (red), Wigner* (orange), and \ac{QT} (blue) -- to the experimental cross-section (grey dashed curve) obtained by combining data from Refs.~\citenum{matthews2005importance} and \citenum{vaghjiani1989absorption}, as recommended in the MPI-Mainz UV/Vis Spectral Atlas.\cite{keller2013mpi}. Note that the experimental photoabsorption cross-section appears smooth and structureless because of the dissociative (\textit{i.e.}, unbound) nature of the potential energy surfaces of the excited electronic states.

The fact that $\sigma (\lambda)$ obtained with a Wigner sampling is closer to the experimental reference than that obtained with \ac{QT} or Wigner* -- both exhibiting a too low overall cross-section -- may appear deceptive at first glance. 
However, the electronic structure method used in this work, namely XMS-CASPT2, exhibits too small oscillator strengths with respect to a reference method like CC3 or even LR-TDDFT (see SI for a detailed benchmark). This observation explains why combining QT sampling with LR-TDDFT for the transition energies and oscillator strengths provided a photoabsorption cross-section in excellent agreement with the experimental one in an earlier work.\cite{prlj2021calculating} The increase of absorption intensities observed for the Wigner-based cross-section (low-energy tail) is an artifact caused by the broad \ce{O-H} bond length distribution provided by this sampling method (see Fig.~\ref{fig:ground-state-distributions}) that causes the transitions of $n'\sigma^\ast$(\ce{O-H}) character (transitions with large oscillator strength) to fall down in energy. These intense transitions pollute the low-energy tail of the spectrum and increase the overall photoabsorption cross-section in this region (as discussed in detail in Ref.~\citenum{prlj2021calculating}). While we regard this effect as artificial, its impact on quantum yields is yet to be determined. Since the nonphysical stretching of the \ce{O-H} bond is absent in \ac{QT} and Wigner* sampling, their cross-sections in the tail region are smaller and smoother, being mainly built from $n\sigma^\ast$(\ce{O-O}) transitions of very weak intensity. Employing a method like CC3 to calculate the transition energies and oscillator strengths on the support of the \ac{QT}- or Wigner*-sampled geometries would lead to a calculated photoabsorption cross-section in better agreement with the experimental one in terms of its intensity for the good reason and not due to an artifact as observed here with the Wigner-sampled geometries.

Focusing now on the higher-energy spectral range\footnote{We stress here that our calculations only include transitions towards the three lowest singlet excited electronic states.}, the differences observed between the results obtained with \ac{QT} and Wigner* indicate that filtering out the problematic low-energy frequency from a standard Wigner sampling does not provide results equivalent to \ac{QT}. Wigner* exhibits a broad (low-intensity) band at around 200~nm that does not appear in the \ac{QT} spectrum. Extending the range of the photoabsorption cross-sections to 150~nm reveals that a broad high-intensity band at 170~nm in the Wigner and QT cross-sections appears much narrower when the Wigner* sampling is used (see Fig.~S3 in the SI). A scan of the potential energy curves along the \ce{C-O-O-H} dihedral angle (see Fig.~S4 in the SI) reveals that the excited electronic states and their transition dipole moments with S$_0$ are significantly affected by this torsion, conversely to the ground state.  
Hence, simply removing the torsion along the \ce{C-O-O-H} dihedral angle from the sampling process, as done with Wigner*, may solve one issue (the artificial \ce{O-H} bond lengths) but lead to an improper account of the role of the torsion in the photoabsorption of \ac{MHP}. In other words, correcting a Wigner sampling by removing low-frequency torsions may be hazardous when these torsions potentially act as photoactive modes.  

\subsection{Wavelength-dependent quantum yields of methylhydroperoxide}

Let us now concentrate on the influence of the \ac{ICs} on the determination of wavelength-dependent quantum yields, $\phi (\lambda)$, calculated from the \ac{TSH} simulations. We start by comparing $\phi (\lambda)$ obtained from an uniform selection of \ac{ICs} sampled from the different distributions -- Wigner, Wigner*, or \ac{QT}. In the uniform selection, we selected \ac{ICs} randomly from the specified distribution within each energy window, without applying any other filters. We will discuss later the results obtained from an $f$-biased sampling of the \ac{ICs}, where the probability of selecting a particular IC is influenced by its oscillator strengths (see, for example, Ref.~\citenum{barbatti2022newton}). 

\ac{MHP} has two main photolysis channels --- the photo-triggered release of a \ce{OH} radical or an H atom. The other minor channels at higher excitation energies involve the photodissociation of an O atom (combined with the formation of methanol) or the simultaneous photodissociation of an H and O atom. The photolysis channel followed by the excited \ac{MHP} molecule is mainly determined by the initial character of the excited electronic state reached by the light-absorption process. For the excitation wavelengths explored in this work, $\phi_{OH}+\phi_{H} \approx 1$ such that we focus on the $\phi_{H}$ for our analysis ($\phi_{H}$ values being more directly related to the issues with ground-state distributions).

The experimental photolysis quantum yields for \ac{MHP} were measured by Vaghjiani et al. at an excitation wavelength of 248~nm, corresponding to the edge of the low-energy tail of $\sigma (\lambda)$ (see Fig.~\ref{fig1}).\cite{vaghjiani1990photodissociation} At this wavelength, $\phi_{H}=0.038 \pm 0.007$ and $\phi_{OH}$ was estimated to be $1.00 \pm 0.18$.
Thelen et al. measured the photodissociation at 193~nm and 248~nm using photofragment translational spectroscopy and did not observe H dissociation (this work does not report quantum yields).\cite{thelen1993photofragmentation}
Blitz et al.\cite{blitz2005wavelength} measured a quantum yield for the \ce{CH3O} radical of $0.81 \pm 0.12$ at 223~nm, which indirectly informs on the value of $\phi_{OH}$ (considering that only OH is formed from the photolysis of MHP at this wavelength). We connect this value to $\phi_{H}$ by $1-\phi_{OH}$, but note that this value should be taken cautiously.

We investigated the wavelength dependence of $\phi_{H}$ by defining three equidistant narrow energy windows centered at 248, 217, and 193~nm. These windows were used to select \ac{ICs} from the three different ground-state sampling strategies. \ac{TSH} simulations were then conducted based on these \ac{ICs}, leading to the prediction of $\phi_{H}$ for each window and each sampling technique (Fig.~\ref{fig1}, left axis). The choice of \ac{ICs} was labeled as uniform, meaning that all \ac{ICs} with vertical transitions falling within the narrow excitation windows were accepted -- the ensuing trajectories were calculated and treated as equally important events. In this way, $\phi_{H}$ was computed as $N_H^{window} / N^{window}$, where $N_H^{window}$ is the total number of trajectories starting from a given excitation window and following the H photodissociation pathway, while $N^{window}$ is the total number of trajectories launched from a window, regardless of their outcome. The error bars were estimated following Ref.~\citenum{persico2014overview}. The \ac{TSH} results obtained from a Wigner sampling (\ac{TSH}/Wigner in the following) differ from that obtained with Wigner* (\ac{TSH}/Wigner*) and \ac{QT} (\ac{TSH}/\ac{QT}) at all wavelengths. Not unexpectedly, the \ac{TSH}/Wigner predicts a larger quantum yield for the H dissociation in the low-energy window. This behavior is directly correlated with the artificially broad \ce{O-H} bond distribution created by a Wigner sampling and discussed in Sec.~\ref{secgeom} and above for the case of photoabsorption cross-sections. \ac{TSH}/Wigner* predicts almost no H dissociation (2 trajectories out of 487), while the \ac{TSH}/\ac{QT} quantum yield for this channel is slightly larger (7 trajectories out of 575). Hence, the $\phi_{H}$ values predicted by \ac{TSH}/\ac{QT} and \ac{TSH}/Wigner are somewhat smaller than the experimental reference (including its error bar), while $\phi_{H}$ predicted with TSH/Wigner is larger. 

Considering that there are electronic transitions with high and low oscillator strengths within a given excitation window, one may wonder whether it is reasonable to assign them the same weight in the process of selecting \ac{ICs}. Since the oscillator strength correlates with the light-absorption probability, one can test whether the nuclear configurations that have larger oscillator strengths should be preferably selected instead of those with low oscillator strengths (within a certain window). 
Such an $f$-biased selection was proposed in the literature\cite{barbatti2022newton} and implemented in \ac{TSH} codes such as Newton-X\cite{barbatti2014newton,barbatti2022newton} and SHARC.\cite{richter2011sharc,mai2018nonadiabatic} 
The $f$-biased selection of \ac{ICs} implies that a given electronic transition, labeled $i$, is associated with a probability calculated as $f_i/f_{max}$, with $f_i$ being the oscillator strength of this transition (alternatively, one can use the Einstein coefficient B, which is proportional to the square of the transition dipole moment). $f_{max}$ corresponds to the most intense transition within the selection window. The probability is then compared to a randomly generated number in the [0,1] interval, and an IC is selected if its probability is larger than the random number. In this work, we used an $f$-biased selection based on oscillator strengths in a modified version of the SHARC code, and we note that a selection based on squared transition dipole moments was also proposed by Persico and Granucci.\cite{persico2014overview} Since the oscillator strength differs from the squared transition dipole moment only by a factor proportional to the excitation energy, the two selection schemes are expected to differ only moderately when the sampling is performed within narrow energy windows.

\begin{figure}[!ht]
\centering
\includegraphics[width=0.9\textwidth]{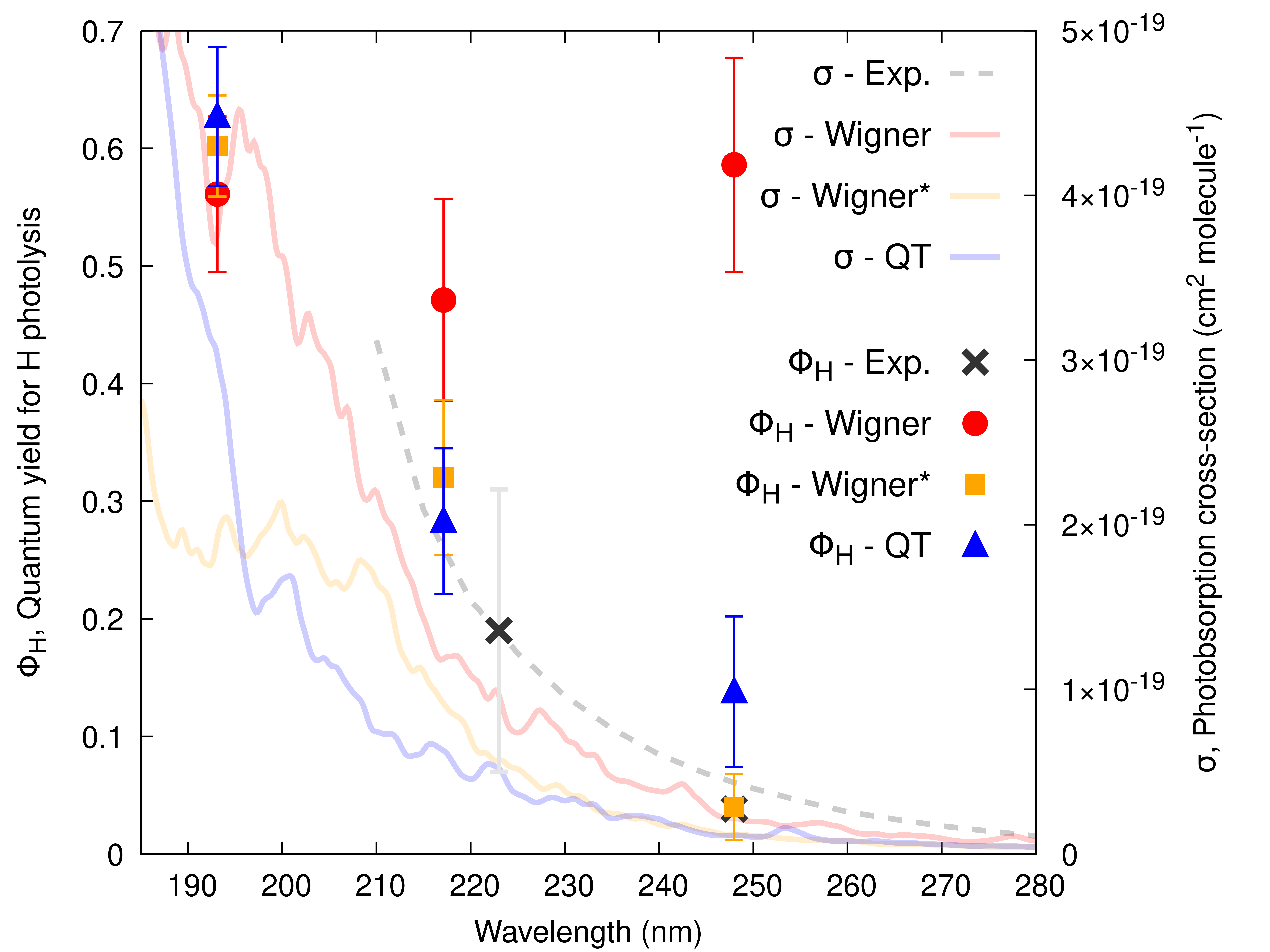}
\caption{Calculated \ac{MHP} photoabsorption cross-sections ($\sigma$) and H dissociation quantum yields ($\phi_H$) obtained from the \ac{NEA} and \ac{TSH} dynamics, respectively, based on an $f$-biased selection of \ac{ICs} (see main text) and compared to experimental data. The same definitions as in Fig.~\ref{fig1} apply for the colors, curves, and symbols.}
\label{fig2}
\end{figure} 

The predicted $\phi_{H}$ obtained from \ac{TSH} simulations initiated with an $f$-biased selection of the \ac{ICs} are presented in Fig.~\ref{fig2} (left axis). The error bars are significantly larger despite using the same pool of \ac{ICs} as for the data in Fig.~\ref{fig1}. The $f$-biased selection would require to calculate a much larger pool of ICs with corresponding excitation energies and oscillator strengths to match the error bars of the uniform selection. The rejection of \ac{ICs} is particularly strong when the energy window contains few very intense transitions and a large number of weak transitions. This behavior can potentially be problematic when transitions with an artificially high oscillator strength appear within the selection window, as we will see.
The $\phi_{H}$ values calculated with the $f$-biased selection are very different from those based on the uniform strategy (compare Fig.~\ref{fig2} to Fig.~\ref{fig1}). As expected from its definition, the $f$-biased selection amplifies the difference between \ac{TSH}/Wigner, \ac{TSH}/Wigner*, and \ac{TSH}/\ac{QT} for \ac{MHP}, reflecting the issue caused by the artificially low-lying $n'\sigma^\ast$(\ce{O-H}) transitions. 
As a result, the $\phi_{H}$ values predicted by \ac{TSH}/Wigner lie around 0.5 across the whole wavelength range. This observation raises a red flag for using an $f$-biased selection strategy when the ground-state sampling affects the balance between bright and dark transitions within a selection window. 
We note that the $f$-biased selection is often set as a default sampling strategy in many standard \ac{TSH} codes. However, the problem does not lie with the $f$-biased selection \textit{per se}, but its combination with an improper sampling technique that amplifies the errors.
\ac{TSH}/Wigner* and \ac{TSH}/\ac{QT} predict $\phi_{H}$ values that are more consistent with available experimental data, although not in perfect agreement.
The $\phi_{H}$ value obtained with \ac{TSH}/\ac{QT} at 248~nm is overestimated, as rare transitions involving brighter $n'\sigma^\ast$(\ce{O-H}) are more likely to be selected than the majority of dark $n\sigma^\ast$(\ce{O-O}) present in this window. The \ac{TSH}/Wigner* dynamics leads to $\phi_{H}$ values that appear closer to the available experimental values. 
Nevertheless, it is difficult to fully assess the $f$-biased selection algorithm without good-quality experimental data over the whole wavelength range.
As noted above, we also need to keep in mind that the electronic-structure method used in this work, XMS-CASPT2, underestimates oscillator strengths, in particular for the lowest $n'\sigma^\ast$(\ce{O-O}) state.

An alternative to the $f$-biased selection strategy would consist in assigning a weight to each \ac{TSH} trajectory initiated from an uniform selection of \ac{ICs}. Effectively, this protocol means that a large number of \ac{TSH} trajectories should be simulated, and the contribution of each TSH trajectory to the calculation of $\phi$ would be weighted by a factor determined from its IC. An earlier work\cite{thompson2018first} proposed that $\phi (\lambda)$ can be calculated as ${\sigma}_{product}(\lambda) / {\sigma}_{tot}(\lambda)$, where ${\sigma}_{product}(\lambda)$ is a photoabsorption cross-section obtained uniquely from the \ac{ICs} that lead to a certain photoproduct, whereas ${\sigma}_{tot}(\lambda)$ is the total cross-section accounting for all \ac{ICs}. Using a ratio of photoabsorption cross-sections is justified by the fact that ${\sigma}(\lambda)$ is proportional to the number of photons absorbed at a wavelength $\lambda$, whereas the number of absorbed photons is proportional to the number of product molecules formed. If we focus on a narrow excitation window and ignore the broadening effects in the \ac{NEA} expression for ${\sigma}$ (see Eq. (2) in Ref.\citenum{prlj2021calculating}), the estimate of $\phi$ reduces to $\sum_i{f_{i,product}^{window}}/\sum_i{f_i^{window}}$, where $\sum_i{f_{i,product}^{window}}$ is a sum over the oscillator strength of the initial conditions $i$ within a window that lead to the formation of a certain product, and $\sum_i{f_i^{window}}$ is the sum over the oscillator strength of all the \ac{ICs} within this energy window. In other words, instead of counting the trajectories yielding a certain photoproduct as done earlier, we may sum up the oscillator strengths of their \ac{ICs} and divide this sum by the total sum of oscillator strengths within the energy window under consideration. Using this strategy with our \ac{NEA} and \ac{TSH} data leads to values for $\phi_{H}$ that are similar to the values obtained from the $f$-biased selection (see Table~S2 in the SI).\footnote{Ultimately, these two schemes should not lead to completely identical results - in the $f$-biased selection strategy, the $f_{i}$ are compared to $f_{max}$ within a given energy window, while in the new scheme discussed here the $f_{i}$ are compared to the \textit{average} $f$ for the energy window of interest.}  This method, however, suffers from an issue in the low-energy tail of the photoabsorption cross-section, where a very small number of trajectories with large initial oscillator strengths (e.g., only 2/487 trajectories for Wigner* at 248 nm) results in relatively high $\phi_{H}$ values (e.g., $0.142$ for Wigner* at 248 nm). The uncertainty for these values is very high and heavily depends on the accuracy of the oscillator strengths employed (a bottleneck for the electronic-structure method used in this work, as discussed in the SI). 

\subsection{Translational kinetic energy distribution for the OH photolysis of methylhydroperoxide}

\begin{figure}[!ht]
\centering
\includegraphics[width=0.8\textwidth]{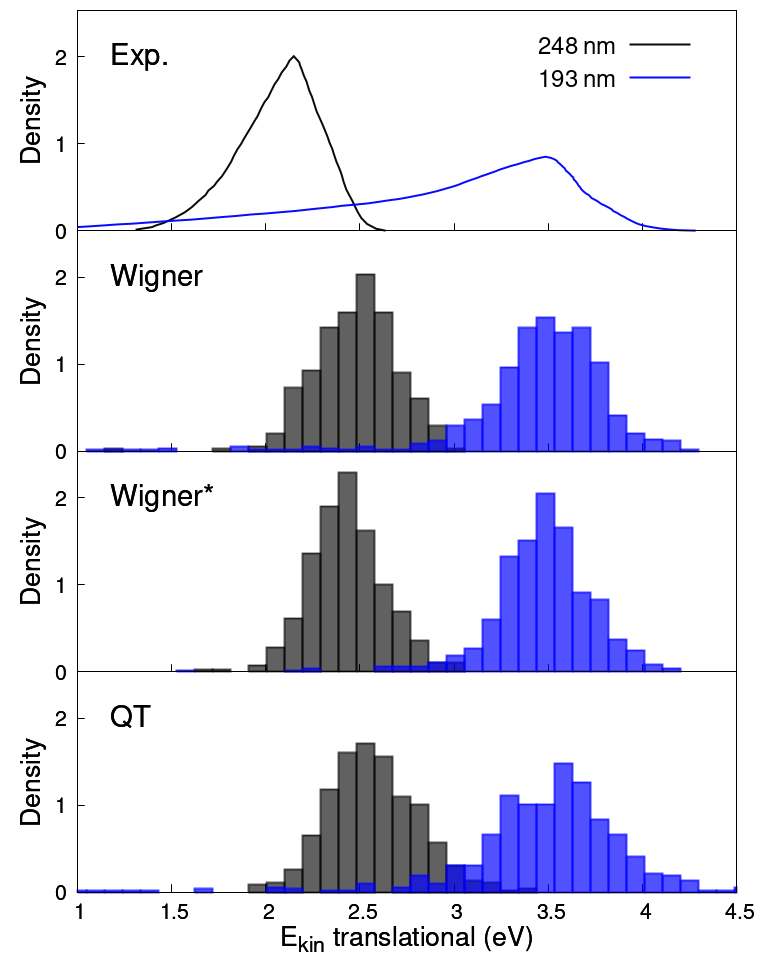}
\caption{Translational kinetic energy maps for OH photodissociation of MHP. Experimental data sets for an excitation at 248 nm (black) and 193~nm (blue) (Ref.~\citenum{thelen1993photofragmentation}) are compared to the theoretical results obtained from \ac{TSH} simulations initiated from a Wigner, Wigner*, or \ac{QT} sampling (with an uniform selection of the \ac{ICs}).}
\label{fig3}
\end{figure}

The final observables considered in this work are translational kinetic energy distributions. Experimentally-derived data for MHP based on measurements in a cold molecular beam are available for \ce{OH} dissociation at 193~nm and 248~nm\cite{thelen1993photofragmentation} and reproduced in the top panel of Fig.~\ref{fig3}. The large swarm of \ac{TSH} trajectories that we generated to calculate $\phi (\lambda)$ allows us to estimate the translational velocities and the kinetic energies of the released \ce{OH} and \ce{CH3O} fragments. 
The results from \ac{TSH}/Wigner, \ac{TSH}/Wigner*, and \ac{TSH}/\ac{QT} samplings with the uniform selection of \ac{ICs} are shown in the lower panels of Fig.~\ref{fig3}.
Overall, the three types of sampling lead to very similar translational kinetic energy maps (Fig.~\ref{fig3}). In all cases, the density peaks are shifted towards higher energies with respect to those observed in the experimental maps. Such a shift can be partly explained by the electronic-structure method employed. XMS-CASPT2/def2-SVPD underestimates the \ce{OH} ground-state dissociation limit by 0.17 eV when compared to UCCSD(F12*)(T)/aug-cc-pVQZ, which is consistent with the higher kinetic energies of the fragment following the photodissociation event (the ground and excited states are near degenerate in the dissociation limit). Also, our \ac{TSH} trajectories are relatively short, and the estimated kinetic energies may not be fully converged for all trajectories, \textit{i.e.}, fragments may still feel a weak interaction at the end of the simulation when the kinetic energy is determined. 

An exciting and counter-intuitive feature of both theoretical and experimental distributions is the long tail at low kinetic energy that appears only for the 193~nm excitation energy. 
The experimental study of Thelen et al.\cite{thelen1993photofragmentation} lacks any explanation of this feature, while the theoretical work of Mahata et al.\cite{mahata2021photodissociation} did not predict the tail. We analyzed the \ac{TSH} trajectories leading to very low translational kinetic energies for the released \ce{OH} and noticed that these \ce{OH} fragments exhibit large vibrational amplitudes. The most common scenario observed for the creation of these vibrationally-excited \ce{OH} fragments is an initial \ce{H} photodissociation, followed by the \ce{O-O} bond cleavage that happens due to a nonadiabatic interaction between the $n\sigma^\ast$(\ce{O-H}) and $n\sigma^\ast$(\ce{O-O}) states. Collision of the departing O and H atoms then creates the \ce{OH} fragment with a high vibrational and low translational energy. Fig~S5 provides an example of such a TSH trajectory.
As these events usually start with an H cleavage, the corresponding initial excitation that could lead to such processes should possess a relatively large oscillator strength.
Hence, we also determined translational kinetic energy maps using a weighted IC selection, which accounts for the initial oscillator strengths of the \ac{ICs}. Since the $f$-biasing scheme used above did not lead to a sufficient number of \ce{OH} trajectories for a meaningful analysis, we employ an \textit{a posteriori} correction by assigning weights to the \ac{TSH} trajectories obtained from the uniform selection.
The translational kinetic energy maps calculated from this biased weighing are shown in the SI, Fig.~S6. 
The differences between the maps obtained with \ac{TSH}/Wigner, \ac{TSH}/Wigner*, and \ac{TSH}/\ac{QT} samplings become more pronounced, with the \ac{TSH}/\ac{QT} results appearing to be the closest to the experimental data. In all cases, the weighted theoretical maps show a further enhanced tail for the excitation at 193~nm leading to a closer agreement with experimental evidence, although a full convergence of these results would require a significantly larger number of \ac{TSH} trajectories. In any case, the results obtained for translational kinetic energy maps appear to advocate further a potential bias of the observables calculated from \ac{TSH} simulations.

\section{Conclusions}
In summary, this work explored how initial conditions and their sampling influence the calculation of photochemical observables when using the nuclear ensemble approach and trajectory surface hopping simulations. As the photochemical quantities determined in this work -- namely photoabsorption cross-sections, wavelength-dependent quantum yields, and translational kinetic energy maps --  are of potential use in atmospheric photochemistry, we use as a test case the photodynamics of the \ac{MHP} molecule, which exhibits a complex electronic structure and challenges standard protocols used in computational photochemistry. The predicted observables appear to depend significantly on the choice of initial conditions, in particular when the approximation underlying a sampling strategy leads to artificial distortions of the molecule along photoactive modes. The impact of the initial conditions on the results of excited-state dynamics simulations highlighted here is not limited to surface hopping simulations but would apply to other mixed quantum/classical\cite{crespo2018recent,doi:10.1021/acs.jctc.5b01180,curchod2018ct} or Gaussian-based methods.\cite{Ben-Nun1998,makhov2014ab,curchod2018ab,lassmann2021aimswiss} 
Despite the limited amount of experimental data available and the approximate electronic structure used in present calculations, the \ac{TSH} dynamics based on a \ac{QT} sampling appears to provide more reliable results than the dynamics initiated from Wigner sampling, though only when oscillator strengths are properly taken into account - either by biasing the selection of \ac{ICs} or equivalently weighing the results at the end of the simulation. The benefit of biasing the selection of \ac{ICs} was spotlighted for calculating the wavelength-dependent quantum yield for H photodissociation and the translational energy maps for OH photodissociation. From a photochemical perspective, the TSH/XMS-CASPT2 simulations presented in this work indicate that the low-energy tail in the translational kinetic energy maps is caused by nonadiabatic processes leading to the formation of a highly vibrationally excited OH fragment. The removal of low-energy normal modes, here a torsion, from the construction of a Wigner distribution leads to improved results for the photoabsorption cross-section at low energy but hampers an adequate description of this quantity at higher energy as the torsion affects high-energy electronic states. Hence, this work advocates a careful evaluation of the approximations underlying a sampling strategy for initial conditions used in excited-state dynamics, in particular when the low-energy modes of a molecule affect the electronic states of interest to its photochemistry.

%%%%%%%%%%%%%%%%%%%%%%%%%%%%%%%%%%%%%%%%%%%%%%%%%%%%%%%%%%%%%%%%%%%%%
%% The "Acknowledgement" section can be given in all manuscript
%% classes.  This should be given within the "acknowledgement"
%% environment, which will make the correct section or running title.
%%%%%%%%%%%%%%%%%%%%%%%%%%%%%%%%%%%%%%%%%%%%%%%%%%%%%%%%%%%%%%%%%%%%%
\begin{acknowledgement}

We would like to thank Prof. Claire Vallance for insightful discussions and Dr. Veronika Jur\'{a}skov\'{a} for helpful suggestions on data visualization.
This project has received funding from the European Research Council (ERC) under the European Union's Horizon 2020 research and innovation programme (Grant agreement No. 803718, project SINDAM) and the EPSRC Grant EP/V026690/1. 
This work made use of the facilities of the Hamilton HPC Service of Durham University.

\end{acknowledgement}

%%%%%%%%%%%%%%%%%%%%%%%%%%%%%%%%%%%%%%%%%%%%%%%%%%%%%%%%%%%%%%%%%%%%%
%% The same is true for Supporting Information, which should use the
%% suppinfo environment.
%%%%%%%%%%%%%%%%%%%%%%%%%%%%%%%%%%%%%%%%%%%%%%%%%%%%%%%%%%%%%%%%%%%%%
\begin{suppinfo}

The Supporting Information contains a detailed benchmark of electronic-structure methods for the low-lying excited states of MHP, a comment on the discarded trajectories from the TSH simulations, tables containing the total number of TSH trajectories simulated for each sampling strategy, the raw data for the quantum yields presented in the main text, and the supporting figures mentioned in the main text. (PDF)

\end{suppinfo}

%%%%%%%%%%%%%%%%%%%%%%%%%%%%%%%%%%%%%%%%%%%%%%%%%%%%%%%%%%%%%%%%%%%%%
%% The appropriate \bibliography command should be placed here.
%% Notice that the class file automatically sets \bibliographystyle
%% and also names the section correctly.
%%%%%%%%%%%%%%%%%%%%%%%%%%%%%%%%%%%%%%%%%%%%%%%%%%%%%%%%%%%%%%%%%%%%%
%\bibliography{peroxide}

\begin{mcitethebibliography}{79}
\providecommand*\natexlab[1]{#1}
\providecommand*\mciteSetBstSublistMode[1]{}
\providecommand*\mciteSetBstMaxWidthForm[2]{}
\providecommand*\mciteBstWouldAddEndPuncttrue
  {\def\EndOfBibitem{\unskip.}}
\providecommand*\mciteBstWouldAddEndPunctfalse
  {\let\EndOfBibitem\relax}
\providecommand*\mciteSetBstMidEndSepPunct[3]{}
\providecommand*\mciteSetBstSublistLabelBeginEnd[3]{}
\providecommand*\EndOfBibitem{}
\mciteSetBstSublistMode{f}
\mciteSetBstMaxWidthForm{subitem}{(\alph{mcitesubitemcount})}
\mciteSetBstSublistLabelBeginEnd
  {\mcitemaxwidthsubitemform\space}
  {\relax}
  {\relax}

\bibitem[Glover \latin{et~al.}(2018)Glover, Mori, Schuurman, Boguslavskiy,
  Schalk, Stolow, and Mart{\'\i}nez]{glover2018excited}
Glover,~W.~J.; Mori,~T.; Schuurman,~M.~S.; Boguslavskiy,~A.~E.; Schalk,~O.;
  Stolow,~A.; Mart{\'\i}nez,~T.~J. Excited state non-adiabatic dynamics of the
  smallest polyene, trans 1,3-butadiene. II. Ab initio multiple spawning
  simulations. \emph{J. Chem. Phys.} \textbf{2018}, \emph{148}, 164303\relax
\mciteBstWouldAddEndPuncttrue
\mciteSetBstMidEndSepPunct{\mcitedefaultmidpunct}
{\mcitedefaultendpunct}{\mcitedefaultseppunct}\relax
\EndOfBibitem
\bibitem[Pathak \latin{et~al.}(2020)Pathak, Ibele, Boll, Callegari, Demidovich,
  Erk, Feifel, Forbes, Di~Fraia, Giannessi, \latin{et~al.}
  others]{pathak2020tracking}
Pathak,~S.; Ibele,~L.~M.; Boll,~R.; Callegari,~C.; Demidovich,~A.; Erk,~B.;
  Feifel,~R.; Forbes,~R.; Di~Fraia,~M.; Giannessi,~L. \latin{et~al.}  Tracking
  the ultraviolet-induced photochemistry of thiophenone during and after
  ultrafast ring opening. \emph{Nat. Chem.} \textbf{2020}, \emph{12},
  795--800\relax
\mciteBstWouldAddEndPuncttrue
\mciteSetBstMidEndSepPunct{\mcitedefaultmidpunct}
{\mcitedefaultendpunct}{\mcitedefaultseppunct}\relax
\EndOfBibitem
\bibitem[Champenois \latin{et~al.}(2021)Champenois, Sanchez, Yang,
  Figueira~Nunes, Attar, Centurion, Forbes, G{\"u}hr, Hegazy, Ji,
  \latin{et~al.} others]{champenois2021conformer}
Champenois,~E.; Sanchez,~D.; Yang,~J.; Figueira~Nunes,~J.; Attar,~A.;
  Centurion,~M.; Forbes,~R.; G{\"u}hr,~M.; Hegazy,~K.; Ji,~F. \latin{et~al.}
  Conformer-specific photochemistry imaged in real space and time.
  \emph{Science} \textbf{2021}, \emph{374}, 178--182\relax
\mciteBstWouldAddEndPuncttrue
\mciteSetBstMidEndSepPunct{\mcitedefaultmidpunct}
{\mcitedefaultendpunct}{\mcitedefaultseppunct}\relax
\EndOfBibitem
\bibitem[Yang \latin{et~al.}(2020)Yang, Zhu, F.~Nunes, Yu, Parrish, Wolf,
  Centurion, G{\"u}hr, Li, Liu, \latin{et~al.} others]{yang2020simultaneous}
Yang,~J.; Zhu,~X.; F.~Nunes,~J.~P.; Yu,~J.~K.; Parrish,~R.~M.; Wolf,~T.~J.;
  Centurion,~M.; G{\"u}hr,~M.; Li,~R.; Liu,~Y. \latin{et~al.}  Simultaneous
  observation of nuclear and electronic dynamics by ultrafast electron
  diffraction. \emph{Science} \textbf{2020}, \emph{368}, 885--889\relax
\mciteBstWouldAddEndPuncttrue
\mciteSetBstMidEndSepPunct{\mcitedefaultmidpunct}
{\mcitedefaultendpunct}{\mcitedefaultseppunct}\relax
\EndOfBibitem
\bibitem[Kirrander \latin{et~al.}(2016)Kirrander, Saita, and
  Shalashilin]{kirrander2016ultrafast}
Kirrander,~A.; Saita,~K.; Shalashilin,~D.~V. Ultrafast X-ray scattering from
  molecules. \emph{J. Chem. Theory. Comput.} \textbf{2016}, \emph{12},
  957--967\relax
\mciteBstWouldAddEndPuncttrue
\mciteSetBstMidEndSepPunct{\mcitedefaultmidpunct}
{\mcitedefaultendpunct}{\mcitedefaultseppunct}\relax
\EndOfBibitem
\bibitem[Li \latin{et~al.}(2017)Li, Inhester, Liekhus-Schmaltz, Curchod,
  Snyder~Jr, Medvedev, Cryan, Osipov, Pabst, Vendrell, Bucksbaum, and
  J.]{li2017ultrafast}
Li,~Z.; Inhester,~L.; Liekhus-Schmaltz,~C.; Curchod,~B.~F.; Snyder~Jr,~J.~W.;
  Medvedev,~N.; Cryan,~J.; Osipov,~T.; Pabst,~S.; Vendrell,~O. \latin{et~al.}
  Ultrafast isomerization in acetylene dication after carbon K-shell
  ionization. \emph{Nat. Commun.} \textbf{2017}, \emph{8}, 453\relax
\mciteBstWouldAddEndPuncttrue
\mciteSetBstMidEndSepPunct{\mcitedefaultmidpunct}
{\mcitedefaultendpunct}{\mcitedefaultseppunct}\relax
\EndOfBibitem
\bibitem[Fu \latin{et~al.}(2011)Fu, Shepler, and Bowman]{fu2011three}
Fu,~B.; Shepler,~B.~C.; Bowman,~J.~M. Three-state trajectory surface hopping
  studies of the photodissociation dynamics of formaldehyde on ab initio
  potential energy surfaces. \emph{J. Am. Chem. Soc.} \textbf{2011},
  \emph{133}, 7957--7968\relax
\mciteBstWouldAddEndPuncttrue
\mciteSetBstMidEndSepPunct{\mcitedefaultmidpunct}
{\mcitedefaultendpunct}{\mcitedefaultseppunct}\relax
\EndOfBibitem
\bibitem[Granucci and Persico(2007)Granucci, and Persico]{granucci2007excited}
Granucci,~G.; Persico,~M. Excited state dynamics with the direct trajectory
  surface hopping method: azobenzene and its derivatives as a case study.
  \emph{Theor. Chem. Acc.} \textbf{2007}, \emph{117}, 1131--1143\relax
\mciteBstWouldAddEndPuncttrue
\mciteSetBstMidEndSepPunct{\mcitedefaultmidpunct}
{\mcitedefaultendpunct}{\mcitedefaultseppunct}\relax
\EndOfBibitem
\bibitem[Thompson and Tapavicza(2018)Thompson, and
  Tapavicza]{thompson2018first}
Thompson,~T.; Tapavicza,~E. First-Principles Prediction of Wavelength-Dependent
  Product Quantum Yields. \emph{J. Phys. Chem. Lett.} \textbf{2018}, \emph{9},
  4758--4764\relax
\mciteBstWouldAddEndPuncttrue
\mciteSetBstMidEndSepPunct{\mcitedefaultmidpunct}
{\mcitedefaultendpunct}{\mcitedefaultseppunct}\relax
\EndOfBibitem
\bibitem[Ravishankara \latin{et~al.}(2015)Ravishankara, Rudich, and
  Pyle]{ravishankara2015}
Ravishankara,~A.~R.; Rudich,~Y.; Pyle,~J.~A. Role of Chemistry in Earth's
  Climate. \emph{Chem. Rev.} \textbf{2015}, \emph{115}, 3679--3681\relax
\mciteBstWouldAddEndPuncttrue
\mciteSetBstMidEndSepPunct{\mcitedefaultmidpunct}
{\mcitedefaultendpunct}{\mcitedefaultseppunct}\relax
\EndOfBibitem
\bibitem[Atkinson and Arey(2003)Atkinson, and Arey]{atkinson2003atmospheric}
Atkinson,~R.; Arey,~J. Atmospheric Degradation of Volatile Organic Compounds.
  \emph{Chem. Rev.} \textbf{2003}, \emph{103}, 4605--4638\relax
\mciteBstWouldAddEndPuncttrue
\mciteSetBstMidEndSepPunct{\mcitedefaultmidpunct}
{\mcitedefaultendpunct}{\mcitedefaultseppunct}\relax
\EndOfBibitem
\bibitem[Vereecken \latin{et~al.}(2015)Vereecken, Glowacki, and
  Pilling]{vereecken2015theoretical}
Vereecken,~L.; Glowacki,~D.~R.; Pilling,~M.~J. Theoretical Chemical Kinetics in
  Tropospheric Chemistry: Methodologies and Applications. \emph{Chem. Rev.}
  \textbf{2015}, \emph{115}, 4063--4114\relax
\mciteBstWouldAddEndPuncttrue
\mciteSetBstMidEndSepPunct{\mcitedefaultmidpunct}
{\mcitedefaultendpunct}{\mcitedefaultseppunct}\relax
\EndOfBibitem
\bibitem[Prlj \latin{et~al.}(2020)Prlj, Ibele, Marsili, and
  Curchod]{prlj2020theoretical}
Prlj,~A.; Ibele,~L.~M.; Marsili,~E.; Curchod,~B. F.~E. On the theoretical
  determination of photolysis properties for atmospheric volatile organic
  compounds. \emph{J. Phys. Chem. Lett.} \textbf{2020}, \emph{11},
  5418--5425\relax
\mciteBstWouldAddEndPuncttrue
\mciteSetBstMidEndSepPunct{\mcitedefaultmidpunct}
{\mcitedefaultendpunct}{\mcitedefaultseppunct}\relax
\EndOfBibitem
\bibitem[Crespo-Otero and Barbatti(2012)Crespo-Otero, and
  Barbatti]{crespo2012spectrum}
Crespo-Otero,~R.; Barbatti,~M. Spectrum simulation and decomposition with
  nuclear ensemble: formal derivation and application to benzene, furan and
  2-phenylfuran. \emph{Theor. Chem. Acc.} \textbf{2012}, \emph{131}, 1237\relax
\mciteBstWouldAddEndPuncttrue
\mciteSetBstMidEndSepPunct{\mcitedefaultmidpunct}
{\mcitedefaultendpunct}{\mcitedefaultseppunct}\relax
\EndOfBibitem
\bibitem[Tully(1990)]{tully1990molecular}
Tully,~J.~C. Molecular dynamics with electronic transitions. \emph{J. Chem.
  Phys.} \textbf{1990}, \emph{93}, 1061--1071\relax
\mciteBstWouldAddEndPuncttrue
\mciteSetBstMidEndSepPunct{\mcitedefaultmidpunct}
{\mcitedefaultendpunct}{\mcitedefaultseppunct}\relax
\EndOfBibitem
\bibitem[Marsili \latin{et~al.}(2022)Marsili, Prlj, and
  Curchod]{marsili2022theoretical}
Marsili,~E.; Prlj,~A.; Curchod,~B.~F. A Theoretical Perspective on the Actinic
  Photochemistry of 2-Hydroperoxypropanal. \emph{J. Phys. Chem. A}
  \textbf{2022}, \emph{126}, 5420--5433\relax
\mciteBstWouldAddEndPuncttrue
\mciteSetBstMidEndSepPunct{\mcitedefaultmidpunct}
{\mcitedefaultendpunct}{\mcitedefaultseppunct}\relax
\EndOfBibitem
\bibitem[Hutton and Curchod(2022)Hutton, and Curchod]{hutton2022photodynamics}
Hutton,~L.; Curchod,~B. F.~E. Photodynamics of Gas-Phase Pyruvic Acid Following
  Light Absorption in the Actinic Region. \emph{ChemPhotoChem} \textbf{2022},
  \emph{6}, e202200151\relax
\mciteBstWouldAddEndPuncttrue
\mciteSetBstMidEndSepPunct{\mcitedefaultmidpunct}
{\mcitedefaultendpunct}{\mcitedefaultseppunct}\relax
\EndOfBibitem
\bibitem[Pereira~Rodrigues \latin{et~al.}(2019)Pereira~Rodrigues, Lopes~de
  Lima, de~Andrade, Ventura, do~Monte, and
  Barbatti]{doi:10.1021/acs.jpca.8b12482}
Pereira~Rodrigues,~G.; Lopes~de Lima,~T.~M.; de~Andrade,~R.~B.; Ventura,~E.;
  do~Monte,~S.~A.; Barbatti,~M. Photoinduced Formation of H-Bonded Ion Pair in
  HCFC-133a. \emph{J. Phys. Chem. A} \textbf{2019}, \emph{123},
  1953--1961\relax
\mciteBstWouldAddEndPuncttrue
\mciteSetBstMidEndSepPunct{\mcitedefaultmidpunct}
{\mcitedefaultendpunct}{\mcitedefaultseppunct}\relax
\EndOfBibitem
\bibitem[McGillen \latin{et~al.}(2017)McGillen, Curchod, Chhantyal-Pun, Beames,
  Watson, Khan, McMahon, Shallcross, and Orr-Ewing]{mcgillen2017criegee}
McGillen,~M.~R.; Curchod,~B. F.~E.; Chhantyal-Pun,~R.; Beames,~J.~M.;
  Watson,~N.; Khan,~M. A.~H.; McMahon,~L.; Shallcross,~D.~E.; Orr-Ewing,~A.~J.
  Criegee intermediate--alcohol reactions, a potential source of functionalized
  hydroperoxides in the atmosphere. \emph{ACS Earth Space Chem.} \textbf{2017},
  \emph{1}, 664--672\relax
\mciteBstWouldAddEndPuncttrue
\mciteSetBstMidEndSepPunct{\mcitedefaultmidpunct}
{\mcitedefaultendpunct}{\mcitedefaultseppunct}\relax
\EndOfBibitem
\bibitem[Wang \latin{et~al.}(2023)Wang, Liu, Zou, Karsili, and
  Lester]{D3CP00207A}
Wang,~G.; Liu,~T.; Zou,~M.; Karsili,~T. N.~V.; Lester,~M.~I. UV
  photodissociation dynamics of the acetone oxide Criegee intermediate:
  experiment and theory. \emph{Phys. Chem. Chem. Phys.} \textbf{2023},
  \emph{25}, 7453--7465\relax
\mciteBstWouldAddEndPuncttrue
\mciteSetBstMidEndSepPunct{\mcitedefaultmidpunct}
{\mcitedefaultendpunct}{\mcitedefaultseppunct}\relax
\EndOfBibitem
\bibitem[McCoy \latin{et~al.}(2021)McCoy, Marchetti, Thodika, and
  Karsili]{mccoy2021simple}
McCoy,~J.~C.; Marchetti,~B.; Thodika,~M.; Karsili,~T.~N. {A Simple and
  Efficient Method for Simulating the Electronic Absorption Spectra of Criegee
  Intermediates: Benchmarking on CH$_2$OO and CH$_3$CHOO}. \emph{J. Phys. Chem.
  A} \textbf{2021}, \emph{125}, 4089--4097\relax
\mciteBstWouldAddEndPuncttrue
\mciteSetBstMidEndSepPunct{\mcitedefaultmidpunct}
{\mcitedefaultendpunct}{\mcitedefaultseppunct}\relax
\EndOfBibitem
\bibitem[Wang \latin{et~al.}(2023)Wang, Liu, Zou, Sojdak, Kozlowski, Karsili,
  and Lester]{doi:10.1021/acs.jpca.2c08025}
Wang,~G.; Liu,~T.; Zou,~M.; Sojdak,~C.~A.; Kozlowski,~M.~C.; Karsili,~T. N.~V.;
  Lester,~M.~I. Electronic Spectroscopy and Dissociation Dynamics of
  Vinyl-Substituted Criegee Intermediates: 2-Butenal Oxide and Comparison with
  Methyl Vinyl Ketone Oxide and Methacrolein Oxide Isomers. \emph{J. Phys.
  Chem. A} \textbf{2023}, \emph{127}, 203--215\relax
\mciteBstWouldAddEndPuncttrue
\mciteSetBstMidEndSepPunct{\mcitedefaultmidpunct}
{\mcitedefaultendpunct}{\mcitedefaultseppunct}\relax
\EndOfBibitem
\bibitem[Franc{\'e}s-Monerris \latin{et~al.}(2020)Franc{\'e}s-Monerris,
  Carmona-Garc{\'\i}a, Acu{\~n}a, D{\'a}valos, Cuevas, Kinnison, Francisco,
  Saiz-Lopez, and Roca-Sanju{\'a}n]{frances2020photodissociation}
Franc{\'e}s-Monerris,~A.; Carmona-Garc{\'\i}a,~J.; Acu{\~n}a,~A.~U.;
  D{\'a}valos,~J.~Z.; Cuevas,~C.~A.; Kinnison,~D.~E.; Francisco,~J.~S.;
  Saiz-Lopez,~A.; Roca-Sanju{\'a}n,~D. Photodissociation mechanisms of major
  mercury (II) species in the atmospheric chemical cycle of mercury.
  \emph{Angew. Chem. Int. Ed.} \textbf{2020}, \emph{59}, 7605--7610\relax
\mciteBstWouldAddEndPuncttrue
\mciteSetBstMidEndSepPunct{\mcitedefaultmidpunct}
{\mcitedefaultendpunct}{\mcitedefaultseppunct}\relax
\EndOfBibitem
\bibitem[Carmona-Garc{\'\i}a \latin{et~al.}(2021)Carmona-Garc{\'\i}a,
  Franc{\'e}s-Monerris, Cuevas, Trabelsi, Saiz-Lopez, Francisco, and
  Roca-Sanju{\'a}n]{carmona2021photochemistry}
Carmona-Garc{\'\i}a,~J.; Franc{\'e}s-Monerris,~A.; Cuevas,~C.~A.; Trabelsi,~T.;
  Saiz-Lopez,~A.; Francisco,~J.~S.; Roca-Sanju{\'a}n,~D. Photochemistry and
  Non-adiabatic Photodynamics of the HOSO Radical. \emph{J. Am. Chem. Soc.}
  \textbf{2021}, \emph{143}, 10836--10841\relax
\mciteBstWouldAddEndPuncttrue
\mciteSetBstMidEndSepPunct{\mcitedefaultmidpunct}
{\mcitedefaultendpunct}{\mcitedefaultseppunct}\relax
\EndOfBibitem
\bibitem[Crespo-Otero and Barbatti(2018)Crespo-Otero, and
  Barbatti]{crespo2018recent}
Crespo-Otero,~R.; Barbatti,~M. Recent Advances and Perspectives on Nonadiabatic
  Mixed Quantum--Classical Dynamics. \emph{Chem. Rev.} \textbf{2018},
  \emph{118}, 7026--7068\relax
\mciteBstWouldAddEndPuncttrue
\mciteSetBstMidEndSepPunct{\mcitedefaultmidpunct}
{\mcitedefaultendpunct}{\mcitedefaultseppunct}\relax
\EndOfBibitem
\bibitem[Barbatti(2020)]{barbatti2020simulation}
Barbatti,~M. {Simulation of Excitation by Sunlight in Mixed Quantum-Classical
  Dynamics}. \emph{J. Chem. Theory Comput.} \textbf{2020}, \emph{16},
  4849--4856\relax
\mciteBstWouldAddEndPuncttrue
\mciteSetBstMidEndSepPunct{\mcitedefaultmidpunct}
{\mcitedefaultendpunct}{\mcitedefaultseppunct}\relax
\EndOfBibitem
\bibitem[Barbatti and Sen(2016)Barbatti, and Sen]{barbatti2016effects}
Barbatti,~M.; Sen,~K. Effects of different initial condition samplings on
  photodynamics and spectrum of pyrrole. \emph{Int. J. Quantum Chem.}
  \textbf{2016}, \emph{116}, 762--771\relax
\mciteBstWouldAddEndPuncttrue
\mciteSetBstMidEndSepPunct{\mcitedefaultmidpunct}
{\mcitedefaultendpunct}{\mcitedefaultseppunct}\relax
\EndOfBibitem
\bibitem[Persico and Granucci(2014)Persico, and Granucci]{persico2014overview}
Persico,~M.; Granucci,~G. An overview of nonadiabatic dynamics simulations
  methods, with focus on the direct approach versus the fitting of potential
  energy surfaces. \emph{Theor. Chem. Acc.} \textbf{2014}, \emph{133},
  1526\relax
\mciteBstWouldAddEndPuncttrue
\mciteSetBstMidEndSepPunct{\mcitedefaultmidpunct}
{\mcitedefaultendpunct}{\mcitedefaultseppunct}\relax
\EndOfBibitem
\bibitem[McCoy(2014)]{mccoy2014role}
McCoy,~A.~B. The role of electrical anharmonicity in the association band in
  the water spectrum. \emph{J. Phys. Chem. B} \textbf{2014}, \emph{118},
  8286--8294\relax
\mciteBstWouldAddEndPuncttrue
\mciteSetBstMidEndSepPunct{\mcitedefaultmidpunct}
{\mcitedefaultendpunct}{\mcitedefaultseppunct}\relax
\EndOfBibitem
\bibitem[Suchan \latin{et~al.}(2018)Suchan, Hollas, Curchod, and
  Slav{\'\i}{\v{c}}ek]{suchan2018importance}
Suchan,~J.; Hollas,~D.; Curchod,~B. F.~E.; Slav{\'\i}{\v{c}}ek,~P. On the
  importance of initial conditions for excited-state dynamics. \emph{Faraday
  Discuss.} \textbf{2018}, \emph{212}, 307--330\relax
\mciteBstWouldAddEndPuncttrue
\mciteSetBstMidEndSepPunct{\mcitedefaultmidpunct}
{\mcitedefaultendpunct}{\mcitedefaultseppunct}\relax
\EndOfBibitem
\bibitem[Mai \latin{et~al.}(2018)Mai, Gattuso, Monari, and
  Gonz{\'a}lez]{mai2018novel}
Mai,~S.; Gattuso,~H.; Monari,~A.; Gonz{\'a}lez,~L. Novel
  molecular-dynamics-based protocols for phase space sampling in complex
  systems. \emph{Front. Chem.} \textbf{2018}, \emph{6}, 495\relax
\mciteBstWouldAddEndPuncttrue
\mciteSetBstMidEndSepPunct{\mcitedefaultmidpunct}
{\mcitedefaultendpunct}{\mcitedefaultseppunct}\relax
\EndOfBibitem
\bibitem[Svoboda \latin{et~al.}(2011)Svoboda, On{\v{c}}{\'a}k, and
  Slav{\'\i}{\v{c}}ek]{svoboda2011simulations}
Svoboda,~O.; On{\v{c}}{\'a}k,~M.; Slav{\'\i}{\v{c}}ek,~P. Simulations of light
  induced processes in water based on ab initio path integrals molecular
  dynamics. I. Photoabsorption. \emph{J. Chem. Phys.} \textbf{2011},
  \emph{135}, 154301\relax
\mciteBstWouldAddEndPuncttrue
\mciteSetBstMidEndSepPunct{\mcitedefaultmidpunct}
{\mcitedefaultendpunct}{\mcitedefaultseppunct}\relax
\EndOfBibitem
\bibitem[Favero \latin{et~al.}(2013)Favero, Granucci, and
  Persico]{favero2013dynamics}
Favero,~L.; Granucci,~G.; Persico,~M. Dynamics of acetone photodissociation: a
  surface hopping study. \emph{Phys. Chem. Chem. Phys.} \textbf{2013},
  \emph{15}, 20651--20661\relax
\mciteBstWouldAddEndPuncttrue
\mciteSetBstMidEndSepPunct{\mcitedefaultmidpunct}
{\mcitedefaultendpunct}{\mcitedefaultseppunct}\relax
\EndOfBibitem
\bibitem[Ceriotti \latin{et~al.}(2009)Ceriotti, Bussi, and
  Parrinello]{ceriotti2009nuclear}
Ceriotti,~M.; Bussi,~G.; Parrinello,~M. Nuclear quantum effects in solids using
  a colored-noise thermostat. \emph{Phys. Rev. Lett.} \textbf{2009},
  \emph{103}, 030603\relax
\mciteBstWouldAddEndPuncttrue
\mciteSetBstMidEndSepPunct{\mcitedefaultmidpunct}
{\mcitedefaultendpunct}{\mcitedefaultseppunct}\relax
\EndOfBibitem
\bibitem[Ceriotti \latin{et~al.}(2010)Ceriotti, Bussi, and
  Parrinello]{ceriotti2010colored}
Ceriotti,~M.; Bussi,~G.; Parrinello,~M. Colored-noise thermostats {\`a} la
  carte. \emph{J. Chem. Theory Comput.} \textbf{2010}, \emph{6},
  1170--1180\relax
\mciteBstWouldAddEndPuncttrue
\mciteSetBstMidEndSepPunct{\mcitedefaultmidpunct}
{\mcitedefaultendpunct}{\mcitedefaultseppunct}\relax
\EndOfBibitem
\bibitem[Huppert \latin{et~al.}(2022)Huppert, Pl{\'e}, Bonella, Depondt, and
  Finocchi]{Finocchi2022}
Huppert,~S.; Pl{\'e},~T.; Bonella,~S.; Depondt,~P.; Finocchi,~F. Simulation of
  Nuclear Quantum Effects in Condensed Matter Systems via Quantum Baths.
  \emph{Appl. Sci.} \textbf{2022}, \emph{12}, 4756\relax
\mciteBstWouldAddEndPuncttrue
\mciteSetBstMidEndSepPunct{\mcitedefaultmidpunct}
{\mcitedefaultendpunct}{\mcitedefaultseppunct}\relax
\EndOfBibitem
\bibitem[Basire \latin{et~al.}(2013)Basire, Borgis, and
  Vuilleumier]{Vuilleumier2013}
Basire,~M.; Borgis,~D.; Vuilleumier,~R. Computing Wigner distributions and time
  correlation functions using the quantum thermal bath method: application to
  proton transfer spectroscopy. \emph{Phys. Chem. Chem. Phys.} \textbf{2013},
  \emph{15}, 12591--12601\relax
\mciteBstWouldAddEndPuncttrue
\mciteSetBstMidEndSepPunct{\mcitedefaultmidpunct}
{\mcitedefaultendpunct}{\mcitedefaultseppunct}\relax
\EndOfBibitem
\bibitem[Dammak \latin{et~al.}(2009)Dammak, Chalopin, Laroche, Hayoun, and
  Greffet]{Dammak2009}
Dammak,~H.; Chalopin,~Y.; Laroche,~M.; Hayoun,~M.; Greffet,~J.-J. Quantum
  Thermal Bath for Molecular Dynamics Simulation. \emph{Phys. Rev. Lett.}
  \textbf{2009}, \emph{103}, 190601\relax
\mciteBstWouldAddEndPuncttrue
\mciteSetBstMidEndSepPunct{\mcitedefaultmidpunct}
{\mcitedefaultendpunct}{\mcitedefaultseppunct}\relax
\EndOfBibitem
\bibitem[Pl{\'e} \latin{et~al.}(2021)Pl{\'e}, Huppert, Finocchi, Depondt, and
  Bonella]{Bonella2021}
Pl{\'e},~T.; Huppert,~S.; Finocchi,~F.; Depondt,~P.; Bonella,~S. Anharmonic
  spectral features via trajectory-based quantum dynamics: A perturbative
  analysis of the interplay between dynamics and sampling. \emph{J. Chem.
  Phys.} \textbf{2021}, \emph{155}, 104108\relax
\mciteBstWouldAddEndPuncttrue
\mciteSetBstMidEndSepPunct{\mcitedefaultmidpunct}
{\mcitedefaultendpunct}{\mcitedefaultseppunct}\relax
\EndOfBibitem
\bibitem[Mauger \latin{et~al.}(2021)Mauger, Pl{\'e}, Lagard{\`e}re, Bonella,
  Mangaud, Piquemal, and Huppert]{Mauger2021}
Mauger,~N.; Pl{\'e},~T.; Lagard{\`e}re,~L.; Bonella,~S.; Mangaud,~{\'E}.;
  Piquemal,~J.-P.; Huppert,~S. Nuclear Quantum Effects in Liquid Water at Near
  Classical Computational Cost Using the Adaptive Quantum Thermal Bath.
  \emph{J. Phys. Chem. Lett.} \textbf{2021}, \emph{12}, 8285--8291\relax
\mciteBstWouldAddEndPuncttrue
\mciteSetBstMidEndSepPunct{\mcitedefaultmidpunct}
{\mcitedefaultendpunct}{\mcitedefaultseppunct}\relax
\EndOfBibitem
\bibitem[Pl{\'e} \latin{et~al.}(2023)Pl{\'e}, Mauger, Adjoua, Inizan,
  Lagard{\`e}re, Huppert, and Piquemal]{Ple2023}
Pl{\'e},~T.; Mauger,~N.; Adjoua,~O.; Inizan,~T.~J.; Lagard{\`e}re,~L.;
  Huppert,~S.; Piquemal,~J.-P. Routine Molecular Dynamics Simulations Including
  Nuclear Quantum Effects: From Force Fields to Machine Learning Potentials.
  \emph{J. Chem. Theory Comput.} \textbf{2023}, \emph{19}, 1432--1445\relax
\mciteBstWouldAddEndPuncttrue
\mciteSetBstMidEndSepPunct{\mcitedefaultmidpunct}
{\mcitedefaultendpunct}{\mcitedefaultseppunct}\relax
\EndOfBibitem
\bibitem[Brieuc \latin{et~al.}(2016)Brieuc, Bronstein, Dammak, Depondt,
  Finocchi, and Hayoun]{Brieuc2016}
Brieuc,~F.; Bronstein,~Y.; Dammak,~H.; Depondt,~P.; Finocchi,~F.; Hayoun,~M.
  Zero-Point Energy Leakage in Quantum Thermal Bath Molecular Dynamics
  Simulations. \emph{J. Chem. Theory Comput.} \textbf{2016}, \emph{12},
  5688--5697\relax
\mciteBstWouldAddEndPuncttrue
\mciteSetBstMidEndSepPunct{\mcitedefaultmidpunct}
{\mcitedefaultendpunct}{\mcitedefaultseppunct}\relax
\EndOfBibitem
\bibitem[Mangaud \latin{et~al.}(2019)Mangaud, Huppert, Pl{\'{e}}, Depondt,
  Bonella, and Finocchi]{Mangaud2019}
Mangaud,~E.; Huppert,~S.; Pl{\'{e}},~T.; Depondt,~P.; Bonella,~S.; Finocchi,~F.
  The Fluctuation--Dissipation Theorem as a Diagnosis and Cure for Zero-Point
  Energy Leakage in Quantum Thermal Bath Simulations. \emph{J. Chem. Theory
  Comput.} \textbf{2019}, \emph{15}, 2863--2880\relax
\mciteBstWouldAddEndPuncttrue
\mciteSetBstMidEndSepPunct{\mcitedefaultmidpunct}
{\mcitedefaultendpunct}{\mcitedefaultseppunct}\relax
\EndOfBibitem
\bibitem[Prlj \latin{et~al.}(2022)Prlj, Marsili, Hutton, Hollas, Shchepanovska,
  Glowacki, Slav{\'\i}{\v{c}}ek, and Curchod]{prlj2021calculating}
Prlj,~A.; Marsili,~E.; Hutton,~L.; Hollas,~D.; Shchepanovska,~D.;
  Glowacki,~D.~R.; Slav{\'\i}{\v{c}}ek,~P.; Curchod,~B. F.~E. Calculating
  Photoabsorption Cross-Sections for Atmospheric Volatile Organic Compounds.
  \emph{ACS Earth Space Chem.} \textbf{2022}, \emph{6}, 207--217\relax
\mciteBstWouldAddEndPuncttrue
\mciteSetBstMidEndSepPunct{\mcitedefaultmidpunct}
{\mcitedefaultendpunct}{\mcitedefaultseppunct}\relax
\EndOfBibitem
\bibitem[Wang \latin{et~al.}(2023)Wang, Zhao, Chan, Yao, Chen, and
  Abbatt]{wang2023organic}
Wang,~S.; Zhao,~Y.; Chan,~A.~W.; Yao,~M.; Chen,~Z.; Abbatt,~J.~P. Organic
  Peroxides in Aerosol: Key Reactive Intermediates for Multiphase Processes in
  the Atmosphere. \emph{Chem. Rev.} \textbf{2023}, \emph{123}, 1635--1679\relax
\mciteBstWouldAddEndPuncttrue
\mciteSetBstMidEndSepPunct{\mcitedefaultmidpunct}
{\mcitedefaultendpunct}{\mcitedefaultseppunct}\relax
\EndOfBibitem
\bibitem[Shiozaki \latin{et~al.}(2011)Shiozaki, Gy{\H{o}}rffy, Celani, and
  Werner]{shiozaki2011communication}
Shiozaki,~T.; Gy{\H{o}}rffy,~W.; Celani,~P.; Werner,~H.-J. Communication:
  Extended multi-state complete active space second-order perturbation theory:
  Energy and nuclear gradients. \emph{J. Chem. Phys.} \textbf{2011},
  \emph{135}, 081106\relax
\mciteBstWouldAddEndPuncttrue
\mciteSetBstMidEndSepPunct{\mcitedefaultmidpunct}
{\mcitedefaultendpunct}{\mcitedefaultseppunct}\relax
\EndOfBibitem
\bibitem[Shiozaki(2018)]{shiozaki2018bagel}
Shiozaki,~T. BAGEL: Brilliantly advanced general electronic-structure library.
  \emph{Wiley Interdiscip. Rev. Comput. Mol. Sci.} \textbf{2018}, \emph{8},
  e1331\relax
\mciteBstWouldAddEndPuncttrue
\mciteSetBstMidEndSepPunct{\mcitedefaultmidpunct}
{\mcitedefaultendpunct}{\mcitedefaultseppunct}\relax
\EndOfBibitem
\bibitem[Rappoport and Furche(2010)Rappoport, and
  Furche]{rappoport2010property}
Rappoport,~D.; Furche,~F. Property-optimized Gaussian basis sets for molecular
  response calculations. \emph{J. Chem. Phys.} \textbf{2010}, \emph{133},
  134105\relax
\mciteBstWouldAddEndPuncttrue
\mciteSetBstMidEndSepPunct{\mcitedefaultmidpunct}
{\mcitedefaultendpunct}{\mcitedefaultseppunct}\relax
\EndOfBibitem
\bibitem[Vlaisavljevich and Shiozaki(2016)Vlaisavljevich, and
  Shiozaki]{vlaisavljevich2016nuclear}
Vlaisavljevich,~B.; Shiozaki,~T. Nuclear Energy Gradients for Internally
  Contracted Complete Active Space Second-Order Perturbation Theory: Multistate
  Extensions. \emph{J. Chem. Theory Comput.} \textbf{2016}, \emph{12},
  3781--3787\relax
\mciteBstWouldAddEndPuncttrue
\mciteSetBstMidEndSepPunct{\mcitedefaultmidpunct}
{\mcitedefaultendpunct}{\mcitedefaultseppunct}\relax
\EndOfBibitem
\bibitem[Polyak \latin{et~al.}(2019)Polyak, Hutton, Crespo-Otero, Barbatti, and
  Knowles]{polyak2019ultrafast}
Polyak,~I.; Hutton,~L.; Crespo-Otero,~R.; Barbatti,~M.; Knowles,~P.~J.
  Ultrafast photoinduced dynamics of 1, 3-cyclohexadiene using XMS-CASPT2
  surface hopping. \emph{Journal of chemical theory and computation}
  \textbf{2019}, \emph{15}, 3929--3940\relax
\mciteBstWouldAddEndPuncttrue
\mciteSetBstMidEndSepPunct{\mcitedefaultmidpunct}
{\mcitedefaultendpunct}{\mcitedefaultseppunct}\relax
\EndOfBibitem
\bibitem[Loos \latin{et~al.}(2018)Loos, Scemama, Blondel, Garniron, Caffarel,
  and Jacquemin]{loos2018mountaineering}
Loos,~P.-F.; Scemama,~A.; Blondel,~A.; Garniron,~Y.; Caffarel,~M.;
  Jacquemin,~D. A mountaineering strategy to excited states: Highly accurate
  reference energies and benchmarks. \emph{J. Chem. Theory Comput.}
  \textbf{2018}, \emph{14}, 4360--4379\relax
\mciteBstWouldAddEndPuncttrue
\mciteSetBstMidEndSepPunct{\mcitedefaultmidpunct}
{\mcitedefaultendpunct}{\mcitedefaultseppunct}\relax
\EndOfBibitem
\bibitem[Folkestad \latin{et~al.}(2020)Folkestad, Kj{\o}nstad, Myhre, Andersen,
  Balbi, Coriani, Giovannini, Goletto, Haugland, Hutcheson, H{\o}yvik, Moitra,
  Paul, Scavino, Skeidsvoll, Tveten, and Koch]{folkestad2020t}
Folkestad,~S.~D.; Kj{\o}nstad,~E.~F.; Myhre,~R.~H.; Andersen,~J.~H.; Balbi,~A.;
  Coriani,~S.; Giovannini,~T.; Goletto,~L.; Haugland,~T.~S.; Hutcheson,~A.
  \latin{et~al.}  eT 1.0: An open source electronic structure program with
  emphasis on coupled cluster and multilevel methods. \emph{J. Chem. Phys.}
  \textbf{2020}, \emph{152}, 184103\relax
\mciteBstWouldAddEndPuncttrue
\mciteSetBstMidEndSepPunct{\mcitedefaultmidpunct}
{\mcitedefaultendpunct}{\mcitedefaultseppunct}\relax
\EndOfBibitem
\bibitem[Humphrey \latin{et~al.}(1996)Humphrey, Dalke, and
  Schulten]{humphrey1996vmd}
Humphrey,~W.; Dalke,~A.; Schulten,~K. VMD: visual molecular dynamics. \emph{J.
  Mol. Graph. Model.} \textbf{1996}, \emph{14}, 33--38\relax
\mciteBstWouldAddEndPuncttrue
\mciteSetBstMidEndSepPunct{\mcitedefaultmidpunct}
{\mcitedefaultendpunct}{\mcitedefaultseppunct}\relax
\EndOfBibitem
\bibitem[Richter \latin{et~al.}(2011)Richter, Marquetand,
  Gonz{\'a}lez-V{\'a}zquez, Sola, and Gonz{\'a}lez]{richter2011sharc}
Richter,~M.; Marquetand,~P.; Gonz{\'a}lez-V{\'a}zquez,~J.; Sola,~I.;
  Gonz{\'a}lez,~L. SHARC: ab Initio Molecular Dynamics with Surface Hopping in
  the Adiabatic Representation Including Arbitrary Couplings. \emph{J. Chem.
  Theory Comput.} \textbf{2011}, \emph{7}, 1253--1258\relax
\mciteBstWouldAddEndPuncttrue
\mciteSetBstMidEndSepPunct{\mcitedefaultmidpunct}
{\mcitedefaultendpunct}{\mcitedefaultseppunct}\relax
\EndOfBibitem
\bibitem[Mai \latin{et~al.}(2018)Mai, Marquetand, and
  Gonz{\'a}lez]{mai2018nonadiabatic}
Mai,~S.; Marquetand,~P.; Gonz{\'a}lez,~L. Nonadiabatic dynamics: The SHARC
  approach. \emph{Wiley Interdiscip. Rev. Comput. Mol. Sci.} \textbf{2018},
  \emph{8}, e1370\relax
\mciteBstWouldAddEndPuncttrue
\mciteSetBstMidEndSepPunct{\mcitedefaultmidpunct}
{\mcitedefaultendpunct}{\mcitedefaultseppunct}\relax
\EndOfBibitem
\bibitem[Furche \latin{et~al.}(2014)Furche, Ahlrichs, H{\"a}ttig, Klopper,
  Sierka, and Weigend]{furche2014turbomole}
Furche,~F.; Ahlrichs,~R.; H{\"a}ttig,~C.; Klopper,~W.; Sierka,~M.; Weigend,~F.
  Turbomole. \emph{Wiley Interdiscip. Rev. Comput. Mol. Sci.} \textbf{2014},
  \emph{4}, 91--100\relax
\mciteBstWouldAddEndPuncttrue
\mciteSetBstMidEndSepPunct{\mcitedefaultmidpunct}
{\mcitedefaultendpunct}{\mcitedefaultseppunct}\relax
\EndOfBibitem
\bibitem[Watts and Francisco(2006)Watts, and Francisco]{watts2006ground}
Watts,~J.~D.; Francisco,~J.~S. Ground and electronically excited states of
  methyl hydroperoxide: Comparison with hydrogen peroxide. \emph{J. Chem.
  Phys.} \textbf{2006}, \emph{125}, 104301\relax
\mciteBstWouldAddEndPuncttrue
\mciteSetBstMidEndSepPunct{\mcitedefaultmidpunct}
{\mcitedefaultendpunct}{\mcitedefaultseppunct}\relax
\EndOfBibitem
\bibitem[Hollas \latin{et~al.}(2021 (date of access: June 2023))Hollas, Suchan,
  On{\v{c}}{\'{a}}k, Svoboda, and Slav{\'{i}}{\v{c}}ek]{abin}
Hollas,~D.; Suchan,~J.; On{\v{c}}{\'{a}}k,~M.; Svoboda,~O.;
  Slav{\'{i}}{\v{c}}ek,~P. {ABIN: source code available at
  https://github.com/PHOTOX/ABIN}. https://doi.org/10.5281/zenodo.1228463, 2021
  (date of access: June 2023)\relax
\mciteBstWouldAddEndPuncttrue
\mciteSetBstMidEndSepPunct{\mcitedefaultmidpunct}
{\mcitedefaultendpunct}{\mcitedefaultseppunct}\relax
\EndOfBibitem
\bibitem[GLE(2021 (date of access: June 2023))]{GLE4MDwebsite}
{GLE4MD website}. http://gle4md.org/, 2021 (date of access: June 2023)\relax
\mciteBstWouldAddEndPuncttrue
\mciteSetBstMidEndSepPunct{\mcitedefaultmidpunct}
{\mcitedefaultendpunct}{\mcitedefaultseppunct}\relax
\EndOfBibitem
\bibitem[Ceriotti \latin{et~al.}(2011)Ceriotti, Manolopoulos, and
  Parrinello]{ceriotti2011}
Ceriotti,~M.; Manolopoulos,~D.~E.; Parrinello,~M. Accelerating the convergence
  of path integral dynamics with a generalized Langevin equation. \emph{J.
  Chem. Phys.} \textbf{2011}, \emph{134}, 084104\relax
\mciteBstWouldAddEndPuncttrue
\mciteSetBstMidEndSepPunct{\mcitedefaultmidpunct}
{\mcitedefaultendpunct}{\mcitedefaultseppunct}\relax
\EndOfBibitem
\bibitem[Barbatti \latin{et~al.}(2014)Barbatti, Ruckenbauer, Plasser, Pittner,
  Granucci, Persico, and Lischka]{barbatti2014newton}
Barbatti,~M.; Ruckenbauer,~M.; Plasser,~F.; Pittner,~J.; Granucci,~G.;
  Persico,~M.; Lischka,~H. Newton-X: a surface-hopping program for nonadiabatic
  molecular dynamics. \emph{Wiley Interdiscip. Rev. Comput. Mol. Sci.}
  \textbf{2014}, \emph{4}, 26--33\relax
\mciteBstWouldAddEndPuncttrue
\mciteSetBstMidEndSepPunct{\mcitedefaultmidpunct}
{\mcitedefaultendpunct}{\mcitedefaultseppunct}\relax
\EndOfBibitem
\bibitem[Barbatti \latin{et~al.}(2022)Barbatti, Bondanza, Crespo-Otero,
  Demoulin, Dral, Granucci, Kossoski, Lischka, Mennucci, Mukherjee,
  \latin{et~al.} others]{barbatti2022newton}
Barbatti,~M.; Bondanza,~M.; Crespo-Otero,~R.; Demoulin,~B.; Dral,~P.~O.;
  Granucci,~G.; Kossoski,~F.; Lischka,~H.; Mennucci,~B.; Mukherjee,~S.
  \latin{et~al.}  Newton-X Platform: New Software Developments for Surface
  Hopping and Nuclear Ensembles. \emph{Journal of Chemical Theory and
  Computation} \textbf{2022}, \emph{18}, 6851--6865\relax
\mciteBstWouldAddEndPuncttrue
\mciteSetBstMidEndSepPunct{\mcitedefaultmidpunct}
{\mcitedefaultendpunct}{\mcitedefaultseppunct}\relax
\EndOfBibitem
\bibitem[Granucci and Persico(2007)Granucci, and Persico]{granucci2007critical}
Granucci,~G.; Persico,~M. Critical appraisal of the fewest switches algorithm
  for surface hopping. \emph{J. Chem. Phys.} \textbf{2007}, \emph{126},
  134114\relax
\mciteBstWouldAddEndPuncttrue
\mciteSetBstMidEndSepPunct{\mcitedefaultmidpunct}
{\mcitedefaultendpunct}{\mcitedefaultseppunct}\relax
\EndOfBibitem
\bibitem[Pieroni \latin{et~al.}(2023)Pieroni, Becuzzi, Creatini, Granucci, and
  Persico]{doi:10.1021/acs.jctc.3c00024}
Pieroni,~C.; Becuzzi,~F.; Creatini,~L.; Granucci,~G.; Persico,~M. Effect of
  Initial Conditions Sampling on Surface Hopping Simulations in the Ultrashort
  and Picosecond Time Range. Azomethane Photodissociation as a Case Study.
  \emph{J. Chem. Theory Comput.} \textbf{2023}, \emph{19}, 2430--2445\relax
\mciteBstWouldAddEndPuncttrue
\mciteSetBstMidEndSepPunct{\mcitedefaultmidpunct}
{\mcitedefaultendpunct}{\mcitedefaultseppunct}\relax
\EndOfBibitem
\bibitem[Scott(1992)]{Scott1992}
Scott,~D.~W. \emph{Multivariate Density Estimation: Theory, Practice, and
  Visualization}; John Wiley \& Sons, Inc., 1992\relax
\mciteBstWouldAddEndPuncttrue
\mciteSetBstMidEndSepPunct{\mcitedefaultmidpunct}
{\mcitedefaultendpunct}{\mcitedefaultseppunct}\relax
\EndOfBibitem
\bibitem[Matthews \latin{et~al.}(2005)Matthews, Sinha, and
  Francisco]{matthews2005importance}
Matthews,~J.; Sinha,~A.; Francisco,~J.~S. The importance of weak absorption
  features in promoting tropospheric radical production. \emph{Proc. Natl.
  Acad. Sci. U.S.A.} \textbf{2005}, \emph{102}, 7449--7452\relax
\mciteBstWouldAddEndPuncttrue
\mciteSetBstMidEndSepPunct{\mcitedefaultmidpunct}
{\mcitedefaultendpunct}{\mcitedefaultseppunct}\relax
\EndOfBibitem
\bibitem[Vaghjiani and Ravishankara(1989)Vaghjiani, and
  Ravishankara]{vaghjiani1989absorption}
Vaghjiani,~G.~L.; Ravishankara,~A.~R. Absorption cross sections of CH$_3$OOH,
  H$_2$O$_2$, and D$_2$O$_2$ vapors between 210 and 365 nm at 297 K. \emph{J.
  Geophys. Res. Atmos.} \textbf{1989}, \emph{94}, 3487--3492\relax
\mciteBstWouldAddEndPuncttrue
\mciteSetBstMidEndSepPunct{\mcitedefaultmidpunct}
{\mcitedefaultendpunct}{\mcitedefaultseppunct}\relax
\EndOfBibitem
\bibitem[Keller-Rudek \latin{et~al.}(2013)Keller-Rudek, Moortgat, Sander, and
  S{\"o}rensen]{keller2013mpi}
Keller-Rudek,~H.; Moortgat,~G.~K.; Sander,~R.; S{\"o}rensen,~R. The MPI-Mainz
  UV/VIS spectral atlas of gaseous molecules of atmospheric interest.
  \emph{Earth Syst. Sci. Data} \textbf{2013}, \emph{5}, 365--373\relax
\mciteBstWouldAddEndPuncttrue
\mciteSetBstMidEndSepPunct{\mcitedefaultmidpunct}
{\mcitedefaultendpunct}{\mcitedefaultseppunct}\relax
\EndOfBibitem
\bibitem[Vaghjiani and Ravishankara(1990)Vaghjiani, and
  Ravishankara]{vaghjiani1990photodissociation}
Vaghjiani,~G.~L.; Ravishankara,~A.~R. Photodissociation of H$_2$O$_2$ and
  CH$_3$OOH at 248 nm and 298 K: Quantum yields for OH, O($^3$P) and H($^2$S).
  \emph{J. Chem. Phys.} \textbf{1990}, \emph{92}, 996--1003\relax
\mciteBstWouldAddEndPuncttrue
\mciteSetBstMidEndSepPunct{\mcitedefaultmidpunct}
{\mcitedefaultendpunct}{\mcitedefaultseppunct}\relax
\EndOfBibitem
\bibitem[Blitz \latin{et~al.}(2005)Blitz, Heard, and
  Pilling]{blitz2005wavelength}
Blitz,~M.~A.; Heard,~D.~E.; Pilling,~M.~J. Wavelength dependent
  photodissociation of CH$_3$OOH: Quantum yields for CH$_3$O and OH, and
  measurement of the OH+CH$_3$OOH rate coefficient. \emph{J. Photochem.
  Photobiol. A: Chemistry} \textbf{2005}, \emph{176}, 107--113\relax
\mciteBstWouldAddEndPuncttrue
\mciteSetBstMidEndSepPunct{\mcitedefaultmidpunct}
{\mcitedefaultendpunct}{\mcitedefaultseppunct}\relax
\EndOfBibitem
\bibitem[Thelen \latin{et~al.}(1993)Thelen, Felder, and
  Huber]{thelen1993photofragmentation}
Thelen,~M.-A.; Felder,~P.; Huber,~J.~R. The photofragmentation of methyl
  hydroperoxide \ce{CH3OOH} at 193 and 248 nm in a cold molecular beam.
  \emph{Chem. Phys. Lett.} \textbf{1993}, \emph{213}, 275--281\relax
\mciteBstWouldAddEndPuncttrue
\mciteSetBstMidEndSepPunct{\mcitedefaultmidpunct}
{\mcitedefaultendpunct}{\mcitedefaultseppunct}\relax
\EndOfBibitem
\bibitem[Mahata and Maiti(2021)Mahata, and Maiti]{mahata2021photodissociation}
Mahata,~P.; Maiti,~B. Photodissociation Dynamics of Methyl Hydroperoxide at 193
  nm: A Trajectory Surface-Hopping Study. \emph{J. Phys. Chem. A}
  \textbf{2021}, \emph{125}, 10321--10329\relax
\mciteBstWouldAddEndPuncttrue
\mciteSetBstMidEndSepPunct{\mcitedefaultmidpunct}
{\mcitedefaultendpunct}{\mcitedefaultseppunct}\relax
\EndOfBibitem
\bibitem[Agostini \latin{et~al.}(2016)Agostini, Min, Abedi, and
  Gross]{doi:10.1021/acs.jctc.5b01180}
Agostini,~F.; Min,~S.~K.; Abedi,~A.; Gross,~E. K.~U. Quantum-Classical
  Nonadiabatic Dynamics: Coupled- vs Independent-Trajectory Methods. \emph{J.
  Chem. Theory Comput.} \textbf{2016}, \emph{12}, 2127--2143\relax
\mciteBstWouldAddEndPuncttrue
\mciteSetBstMidEndSepPunct{\mcitedefaultmidpunct}
{\mcitedefaultendpunct}{\mcitedefaultseppunct}\relax
\EndOfBibitem
\bibitem[Curchod \latin{et~al.}(2018)Curchod, Agostini, and
  Tavernelli]{curchod2018ct}
Curchod,~B.~F.; Agostini,~F.; Tavernelli,~I. CT-MQC--a coupled-trajectory mixed
  quantum/classical method including nonadiabatic quantum coherence effects.
  \emph{Eur. Phys. J. B} \textbf{2018}, \emph{91}, 1--12\relax
\mciteBstWouldAddEndPuncttrue
\mciteSetBstMidEndSepPunct{\mcitedefaultmidpunct}
{\mcitedefaultendpunct}{\mcitedefaultseppunct}\relax
\EndOfBibitem
\bibitem[Ben-Nun and Mart\'{\i}nez(1998)Ben-Nun, and
  Mart\'{\i}nez]{Ben-Nun1998}
Ben-Nun,~M.; Mart\'{\i}nez,~T.~J. Nonadiabatic molecular dynamics: Validation
  of the multiple spawning method for a multidimensional problem. \emph{J.
  Chem. Phys.} \textbf{1998}, \emph{108}, 7244--7257\relax
\mciteBstWouldAddEndPuncttrue
\mciteSetBstMidEndSepPunct{\mcitedefaultmidpunct}
{\mcitedefaultendpunct}{\mcitedefaultseppunct}\relax
\EndOfBibitem
\bibitem[Makhov \latin{et~al.}(2014)Makhov, Glover, Martinez, and
  Shalashilin]{makhov2014ab}
Makhov,~D.~V.; Glover,~W.~J.; Martinez,~R.~J.; Shalashilin,~D.~V. Ab initio
  multiple cloning algorithm for quantum nonadiabatic molecular dynamics.
  \emph{J. Chem. Phys.} \textbf{2014}, \emph{141}, 054110\relax
\mciteBstWouldAddEndPuncttrue
\mciteSetBstMidEndSepPunct{\mcitedefaultmidpunct}
{\mcitedefaultendpunct}{\mcitedefaultseppunct}\relax
\EndOfBibitem
\bibitem[Curchod and Mart{\'\i}nez(2018)Curchod, and
  Mart{\'\i}nez]{curchod2018ab}
Curchod,~B. F.~E.; Mart{\'\i}nez,~T.~J. Ab initio nonadiabatic quantum
  molecular dynamics. \emph{Chem. Rev.} \textbf{2018}, \emph{118},
  3305--3336\relax
\mciteBstWouldAddEndPuncttrue
\mciteSetBstMidEndSepPunct{\mcitedefaultmidpunct}
{\mcitedefaultendpunct}{\mcitedefaultseppunct}\relax
\EndOfBibitem
\bibitem[Lassmann and Curchod(2021)Lassmann, and Curchod]{lassmann2021aimswiss}
Lassmann,~Y.; Curchod,~B. F.~E. AIMSWISS---Ab initio multiple spawning with
  informed stochastic selections. \emph{J. Chem. Phys.} \textbf{2021},
  \emph{154}, 211106\relax
\mciteBstWouldAddEndPuncttrue
\mciteSetBstMidEndSepPunct{\mcitedefaultmidpunct}
{\mcitedefaultendpunct}{\mcitedefaultseppunct}\relax
\EndOfBibitem
\end{mcitethebibliography}
\providecommand{\latin}[1]{#1}
\makeatletter
\providecommand{\doi}
  {\begingroup\let\do\@makeother\dospecials
  \catcode`\{=1 \catcode`\}=2 \doi@aux}
\providecommand{\doi@aux}[1]{\endgroup\texttt{#1}}
\makeatother
\providecommand*\mcitethebibliography{\thebibliography}
\csname @ifundefined\endcsname{endmcitethebibliography}
  {\let\endmcitethebibliography\endthebibliography}{}

\end{document}

\tableofcontents

\beginsupplement

\section{Additional computational details}

\subsection{Electronic-structure benchmarks}
We benchmarked the reliability of the XMS(4)-CASPT2(8/6)/def2-SVPD level of theory (see Table~\ref{tab:benchmark} below) by comparing the predicted vertical transition energies and oscillator strengths with those obtained with a larger aug-cc-pVTZ\cite{woon1993gaussian} basis set, and against the reference CC3/aug-cc-pVTZ values. We have also tested other single-reference methods of various accuracy combined with basis sets of variable size. A discussion of the results can be found in Sec.~\ref{benchmarkes} below.

Excited-state coupled cluster (CC) calculations~\cite{koch1995excitation} were performed within the equation-of-motion (EOM) formalism in a hierarchy comprising CC2, EOM-CCSD, and CC3.\cite{paul2020new} All CC calculations were performed with the eT v1.8 program package.\cite{folkestad2020t}

Linear-response time-dependent density functional theory (LR-TDDFT)~\cite{runge1984density,casida1995time} was used with the PBE0 exchange-correlation functional~\cite{adamo1999toward} within the Tamm-Dancoff approximation (TDA).~\cite{hirata1999time} Calculations using ADC(2) (algebraic diagrammatic construction up to second order)~\cite{trofimov1995efficient,dreuw2015algebraic} and its spin-component scaled variant (SCS-ADC(2))~\cite{hellweg2008benchmarking} were performed with frozen core and the resolution of the identity approximation~\cite{weigend2002efficient}, using default SCS scaling factors (cos= 1.2, css= 0.33333). LR-TDDFT and ADC(2) excited-state calculations were done with Turbomole 7.4.1.~\cite{furche2014turbomole} 

Turbomole was also employed for the ground-state estimate of the \ce{O-O} dissociation limit. To determine this quantity, relaxed scan was performed with unrestricted Møller–Plesset perturbation theory up to second order (UMP2/aug-cc-pVDZ), while the energies were calculated using explicitly-correlated unrestricted coupled-cluster (CC) singles and doubles with perturbative triples, namely UCCSD(F12*)(T)/aug-cc-pVQZ.\cite{hattig2010communications}

\subsection{Discarded ICs and Trajectories}
A small fraction of ICs and TSH trajectories experienced numerical problems and their potential impact on calculated observables needs to be discussed.  

Out of 4000 ICs from each type of sampling strategies,  81 ICs from Wigner sampling, 58 ICs from Wigner* sampling, and 136 ICs from QT sampling had problems with the convergence of oscillator strengths for either of the three excited states considered. These ICs were not used for the calculations of photoabsorption cross-sections or the TSH dynamics simulations. Considering that their number is relatively small and only an even smaller fraction of these ICs would have fit the narrow energy windows used for initiating TSH trajectories, we do not expect any large impact on calculated observables. We acknowledge that a small number of discarded ICs could potentially affect the fraction of H dissociation in the 248~nm window, since $\Phi_H$(248~nm) depends on rare events. Nevertheless, after checking the discarded ICs that have transitions fitting in this window, we could not find any indication (based on excited-state characters and oscillator strengths) that they would lead to an excess H dissociation.

A fraction of TSH trajectories was discarded due to nonphysical discontinuities in total and/or electronic (potential) energies along the dynamics. Such issues originate from the multireference electronic structure method used (e.g., changes in active space along the trajectory). The number of discarded trajectories for each set of TSH simulations is given in Table~S1. 
The number of discarded trajectories is negligible for the lowest energy window at 248~nm, while it amounts to around 10--15\,\% for the highest window at 193~nm. 
These trajectories were not used in the calculation of the quantum yields and velocity maps. In general, trajectories with total classical energy jumps larger than 15\% of the current kinetic energy and nonphysical steps in electronic energies were discarded. 
However, if the discontinuities occurred after the outcome of photolysis was unambiguous (typically already after 10--15~fs of dynamics), trajectories were still accepted and used in the statistics of the quantum yields. Nevertheless, these trajectories were not used for the determination of the OH kinetic energy maps since OH velocities were collected at 25 fs. The number of OH trajectories with numerical problems occurring after the OH dissociation (but before 25 fs) was between 8 - 18\% for all windows and all types of sampling, implying that their impact on kinetic energy maps should not be decisive in any case. 

\begin{table}[ht]
\caption{Summary of the outcomes of the TSH simulations.}
\label{tab:tsh_summary}
\begin{tabular}{@{}llll|lll@{}}
\toprule
  & \multicolumn{3}{c|}{\textbf{Wigner (uniform)}} & \multicolumn{3}{c}{\textbf{Wigner ($f$-biased)}} \\ \midrule
window: & 248\,nm & 217\,nm & 193\,nm & 248\,nm & 217\,nm & 193\,nm\\ \midrule
OH & 440 & 815 & 675  & 12 & 18 & 24  \\
H  & 34  & 88 & 259 & 17  & 16 & 32 \\
O  & -  & - & 1 & -  & - & 1  \\ 
O+H & - & 1 & 9 & - & - & -  \\
discarded & 4 & 21 & 109 & 1 & 2 & 4\\ 
total    & 478 & 925 & 1053  & 30 & 36 & 61 \\ \midrule
  & \multicolumn{3}{c|}{\textbf{Wigner* (uniform)}} & \multicolumn{3}{c}{\textbf{Wigner* ($f$-biased)}} \\ \midrule
window: & 248\,nm & 217\,nm & 193\,nm & 248\,nm & 217\,nm & 193\,nm\\ \midrule
OH & 485 & 890 & 560 & 48 & 34 & 51   \\
H  & 2  & 26 & 140 & 2  & 16 & 75  \\
O  & -  & - & -  & -  & - & -  \\ 
O+H & - & 1 & 5 & - & - & 2 \\
discarded & - & 18 & 112 & - & 1 & 14 \\ 
total    & 487 & 935 & 817 & 50 & 51 & 142 \\ \midrule
  & \multicolumn{3}{c|}{\textbf{QT (uniform)}} & \multicolumn{3}{c}{\textbf{QT ($f$-biased)}} \\ \midrule
window: & 248\,nm & 217\,nm & 193\,nm & 248\,nm & 217\,nm & 193\,nm\\ \midrule
OH & 568 & 732 & 588 & 25 & 38 & 25   \\
H  & 7  & 38 & 291 & 4  & 15 & 42  \\
O  & -  & - & -  & -  & - & -   \\ 
O+H & - & - & 4  & - & - & - \\
discarded & 6 & 24 & 124  & 1 & 2 & 12 \\ 
total    & 581 & 794 & 1007  & 30 & 55 & 79  \\
 \bottomrule
\end{tabular}
\end{table}

\begin{table}
  \caption{Photolysis quantum yields $\phi_H$ and the corresponding standard deviations as plotted in Fig.~3 and 4 in the main text. Note that $\phi_H$ values include a small fraction of H from the O+H dissociation channel.}
  \label{tbl:table5}
  \begin{tabular}{lcccc}

\toprule
  & \multicolumn{3}{c}{\textbf{Uniform selection}} \\ \midrule
  & 248\,nm & 217\,nm & 193\,nm\\ \midrule
Wigner & 0.072 $\pm$ 0.012 & 0.098 $\pm$ 0.010 & 0.284 $\pm$ 0.015  \\
Wigner* & 0.004 $\pm$ 0.003  & 0.029 $\pm$ 0.006 & 0.206 $\pm$ 0.015 \\
QT  & 0.012 $\pm$ 0.005 & 0.049 $\pm$ 0.008 & 0.334 $\pm$ 0.016   \\ \midrule
  & \multicolumn{3}{c}{\textbf{$f$-biased selection}} \\ \midrule
  & 248\,nm & 217\,nm & 193\,nm\\ \midrule
Wigner & 0.586 $\pm$ 0.091 & 0.471 $\pm$ 0.086 & 0.561 $\pm$ 0.066  \\
Wigner* & 0.040 $\pm$ 0.028  & 0.320 $\pm$ 0.066 & 0.602 $\pm$ 0.043 \\
QT  & 0.138 $\pm$ 0.064  & 0.283 $\pm$ 0.062 & 0.627 $\pm$ 0.059   \\ \midrule
  & \multicolumn{3}{c}{\textbf{A posteriori scaling}} \\ \midrule
  & 248\,nm & 217\,nm & 193\,nm\\ \midrule
Wigner & 0.631 & 0.506 & 0.564  \\
Wigner* & 0.142  & 0.312 & 0.603 \\
QT  & 0.301  & 0.314 & 0.569   \\

    \hline
   
  \end{tabular}
 
\end{table}
\clearpage
\section{Benchmarking electronic-structure methods for the description of MHP excited electronic states}
\label{benchmarkes}

To assess the accuracy of the XMS(4)-CASPT2(8/6)/def2-SVPD level of theory that was used for our NEA and TSH simulations, we benchmarked the excitation energies and oscillator strengths at the Franck-Condon geometry, optimized using the MP2/aug-cc-pVDZ. The results are summarized in Table~\ref{tab:benchmark} and discussed below.

\begin{table}[h!]
\caption{Excitation energies and corresponding oscillator strengths (in parenthesis) for the first three excited singlet electronic states of MHP calculated with different electronic-structure methods. Energies are given in eV.}
\label{tab:benchmark}
\begin{tabular}{@{}llllll@{}}
\toprule
                         & $n'\sigma^\ast$(\ce{O-O}) & $n\sigma^\ast$(\ce{O-O}) & $n'\sigma^\ast$(\ce{O-H}) \\ \midrule
\textbf{EOM-CC3/aug-cc-pVTZ} & 5.66 (0.00133) & 6.93 (0.00461) & 6.81 (0.01060)    \\
EOM-CC3/cc-pVDZ          & 5.88 (0.00032)  & 7.17 (0.00218) & 8.01 (0.01630)  \\
EOM-CC3/aug-cc-pVDZ      & 5.64 (0.00161)  & 6.91 (0.00519) & 6.71 (0.01063)  \\ \midrule
XMS-CASPT2(8/6)/def2-SVPD & 5.72 (0.00019) & 7.21 (0.00024) & 7.23 (0.01046) \\
XMS-CASPT2(8/6)/aug-cc-pVTZ & 5.52 (0.00035) & 7.02 (0.00153) & 6.84 (0.00978) \\
XMS-CASPT2(8/6)/def2-SVPD & 5.66 (0.00033) & 7.11 (0.00108) & 7.04 (0.01065) \\ 
(real vertical shift set to 0.3 a.u.) & & & \\ \midrule
EOM-CCSD/aug-cc-pVDZ     & 5.70 (0.00184) & 7.02 (0.00612) & 6.67 (0.01014) \\
EOM-CCSD/aug-cc-pVTZ     & 5.74 (0.00138) & 7.05 (0.00554) & 6.86 (0.01026) \\ \midrule
EOM-CC2/aug-cc-pVDZ          & 5.64 (0.00439) & 6.97 (0.01018) & 6.13 (0.00800) \\
EOM-CC2/aug-cc-pVTZ          & 5.68 (0.00314) & 7.01 (0.00933)& 6.27 (0.00849)  \\ \midrule
ADC(2)/aug-cc-pVTZ          & 5.69 (0.00230) &  7.01 (0.00828)  & 6.24 (0.00655) \\
SCS-ADC(2)/aug-cc-pVTZ      & 5.90 (0.00164) & 7.21 (0.00728) & 6.66 (0.00820)  \\ \midrule
LR-TDDFT/TDA/PBE0/aug-cc-pVDZ   & 5.47 (0.0009)  & 6.81 (0.0064) & 6.46 (0.0123) \\ \bottomrule
\end{tabular}
\end{table}

\subsection{Reference CC3 results}

The CC3 method with an aug-cc-pVTZ basis set was used as our reference, as CC3 was previously shown to produce highly accurate results for excitation energies and oscillator strengths for a large number of molecular systems.\cite{loos2018mountaineering}.
Comparing the CC3 results for aug-cc-pVDZ and aug-cc-pVTZ allows us to assess the convergence of the results with respect to the basis set, with the largest difference for the electronic energy being 0.1~eV for the $n'\sigma^\ast$(\ce{O-H}) state. 
The partial Rydberg character of this electronic state is revealed by comparing results obtained from a basis set without diffuse functions, cc-pVDZ, leading to a 1.3~eV shift in energy for this electronic transition. The partial Rydberg character of this electronic state is consistent with the conclusions from Ref.~\citenum{prlj2020theoretical} concerning the electronic excitations of \textit{tert}-butylhydroperoxide.

We note that there is also a pronounced basis set dependence of the oscillator strengths of the valence $n\sigma^\ast$(\ce{O-O}) and $n'\sigma^\ast$(\ce{O-O}) states. 
Increasing the basis set size from aug-cc-pVDZ to aug-cc-pVTZ decreases the oscillator strength by 25\%. 
It is thus possible that there is a small residual error for oscillator strengths of these states even with the triple-zeta basis.

\subsection{Performance of XMS-CASPT2}

In our TSH simulations using XMS-CASPT2, we used a relatively modest def2-SVPD basis set. This choice has been dictated by the compromise between accuracy of the electronic-structure method and its computational cost. In terms of the excited-state energies, using an aug-cc-pVTZ basis set with XMS-CASPT2 yields transition energies that are closer to the CC3 reference than XMS-CASPT2/def2-SVPD. 
With def2-SVPD, the excitation energies are within 0.1~eV of the reference for the $n\sigma^\ast$(\ce{O-O}) states, but overestimated by 0.4~eV for the $n'\sigma^\ast$(\ce{O-H}) state. 
These results offer a validation for the XMS-CASPT2 strategy employed in our work, even if our results could have been improved by using a larger basis set -- the cost of such calculations are currently too high to be combined with on-the-fly TSH dynamics. 

A very important observation from our benchmark is that XMS-CASPT2 severely underestimates the oscillator strengths of the $n\sigma^\ast$(\ce{O-O}) and $n'\sigma^\ast$(\ce{O-O})  states, even with aug-cc-pVTZ. In principle, our TSH simulations should be insensitive to the absolute values of the oscillator strengths as long as the deviations are uniform for all electronic state considered. This is unfortunately not the case for MHP --- oscillator strengths of the $n\sigma^\ast$(\ce{O-O}) and $n'\sigma^\ast$(\ce{O-O}) states are severely underestimated, even more so with the smaller def2-SVPD basis, while the (much larger) oscillator strength of $n'\sigma^\ast$(\ce{O-H}) is very close to the reference. When it comes to the calculation of the photoabsorption cross-section for MHP, the severe underestimation of the oscillator strength for the $n\sigma^\ast$(\ce{O-O}) and $n'\sigma^\ast$(\ce{O-O}) transitions leads to a too low cross-section in the low-energy tail, as observed and discussed in the main text.

\subsection{Accuracy of single-reference methods and LR-TDDFT}

Among all the other electronic-structure methods tested, only EOM-CCSD was able to provide quantitatively consistent results for both energies and oscillator strengths across all excited states considered. Both CC2 and ADC2 severely underestimate the energy of the $n'\sigma^\ast$(\ce{O-H}) transition. The inaccuracy of CC2 and ADC(2) for Rydberg states has been reported in the literature.\cite{Szalay2017} This limitation can be alleviated by using the SCS scheme (as in SCS-ADC(2)), at the cost of deteriorating the description of the electronic energies for the pure valence states (by up to 0.2~eV). This observation is again consistent with previous findings, see for example Table~1 in Ref.~\citenum{Szalay2020}.

Interestingly, CC2 (and to a lesser extent ADC(2)) overestimates the oscillator strengths of the $n\sigma^\ast$(\ce{O-O}) and $n'\sigma^\ast$(\ce{O-O}) transitions by a factor of two, even though the excitation energies of these states are close to the reference results. This observation again contrasts with the significant underestimation of these quantities by XMS-CASPT2.

We have also tested LR-TDDFT/TDA using the PBE0 functional. Similarly to CC2 and ADC2, the energy of the $n'\sigma^\ast$(\ce{O-H}) transition with the partial Rydberg character is underestimated, while the energies of the other two electronic states are within 0.2~eV from the reference. Overall, the oscillator strengths obtained with LR-TDDFT/TDA/PBE0 show a more balanced agreement with the CC3 reference values than XMS-CASPT2. 

\newpage

\section{Supplementary figures}

\begin{figure}[!ht]
\centering
\includegraphics[width=1.0\textwidth]{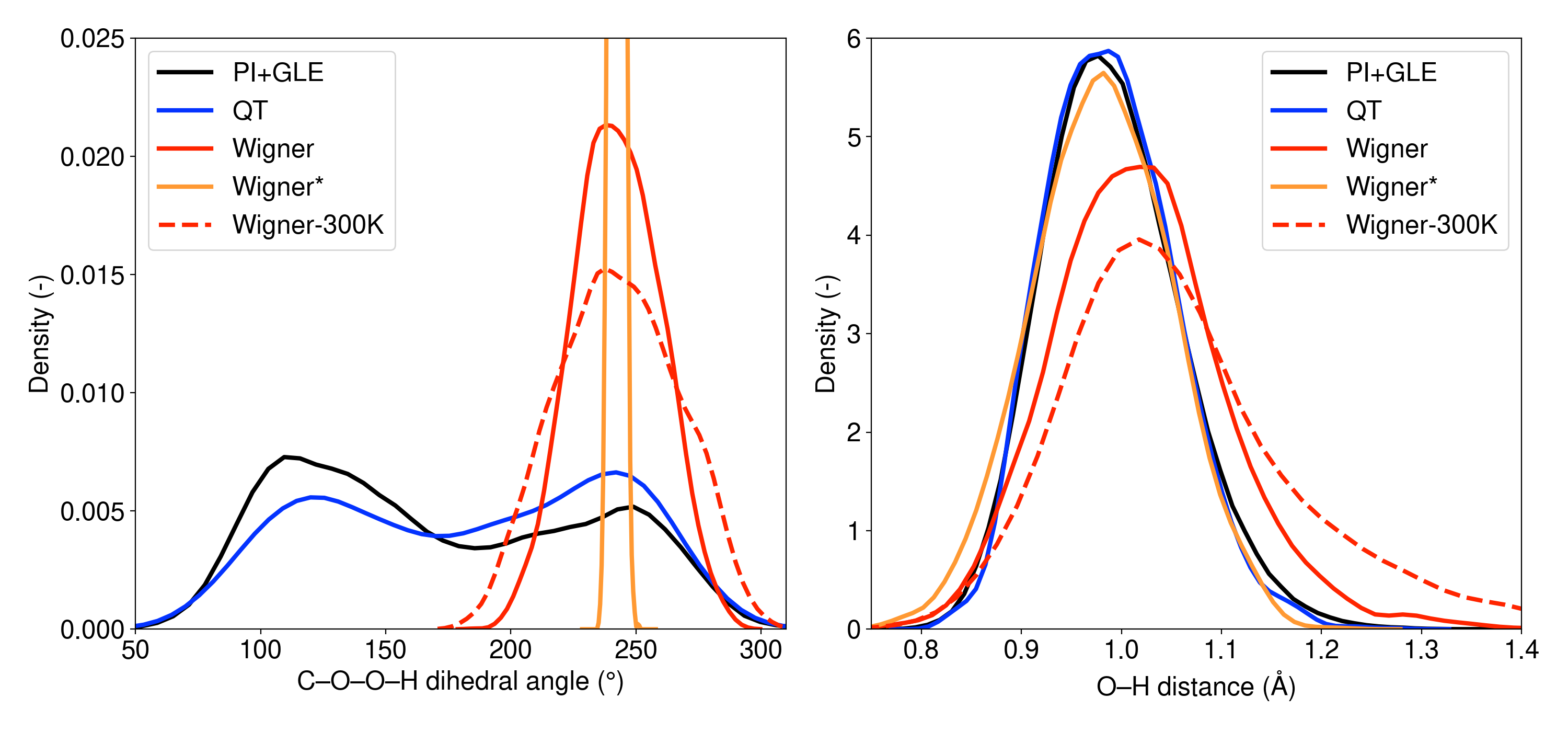}
\caption{Ground-state density distribution for the \ce{C-O-O-H} dihedral angle (left panel) and \ce{O-H} distance (right panel) of MHP obtained from different sampling procedures, namely Wigner, Wigner*, QT, PI+GLE, and Wigner at 300\,K (Wigner-300K). The data presented for the QT, Wigner, and Wigner* distributions correspond to the data shown in Fig.~2 from the main text. The distributions were smoothed using a Gaussian kernel density estimation and the kernel bandwidth was estimated using Scott's rule.\cite{Scott1992}}
\label{fig:coordinate_sampling}
\end{figure}

\begin{figure}[!ht]
\centering
\includegraphics[width=0.8\textwidth]{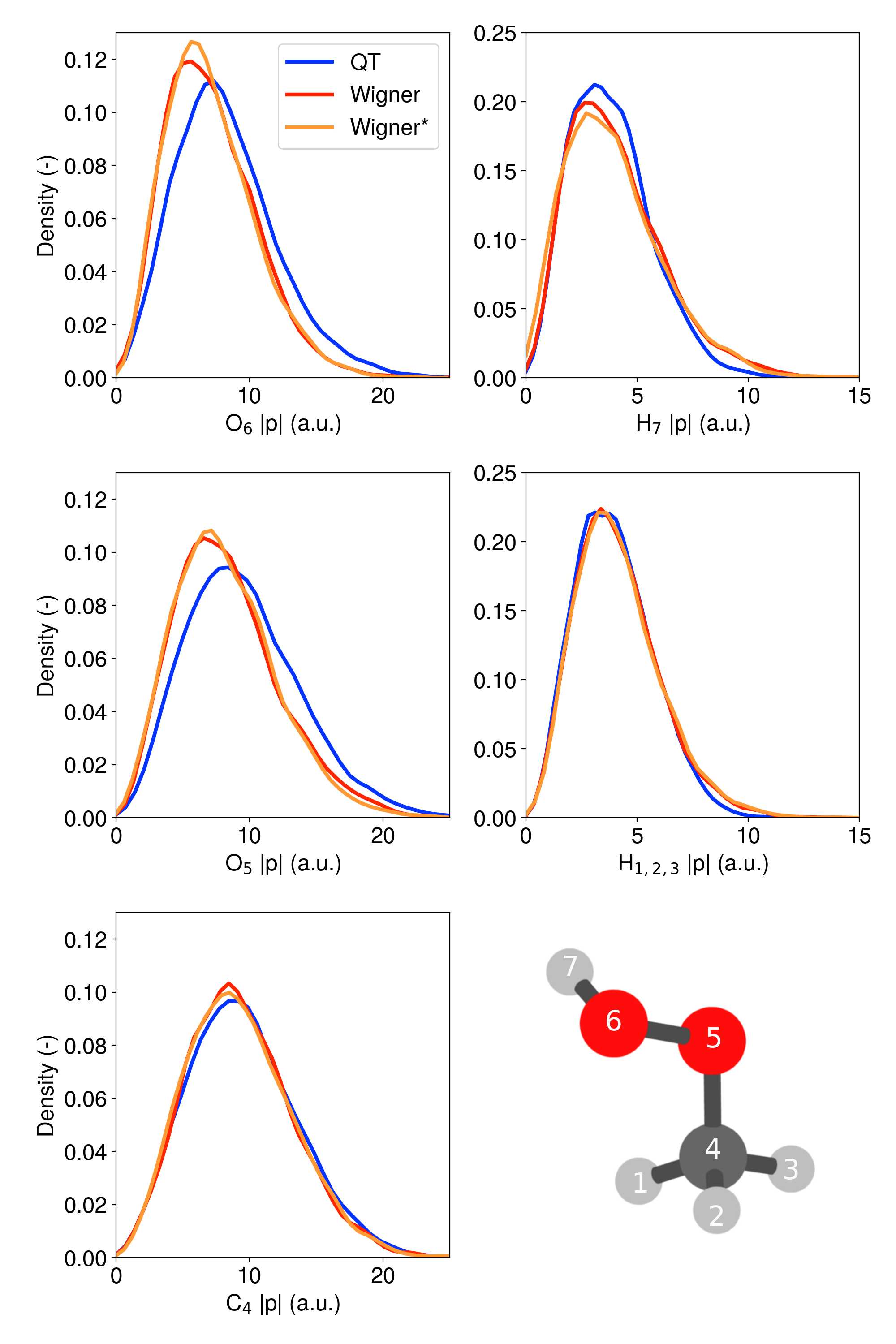}
\caption{Nuclear momentum distributions obtained from different approximate ground-state sampling strategies. The plots were obtained by analyzing the nuclear momenta for the different sets of 4000 ICs discussed in the main text. The distributions were smoothed using a Gaussian kernel density estimation and the kernel bandwidth was estimated using Scott's rule.\cite{Scott1992}}
\label{fig:momentum_sampling}
\end{figure}

\begin{figure}[!ht]
\centering
\includegraphics[width=0.8\textwidth]{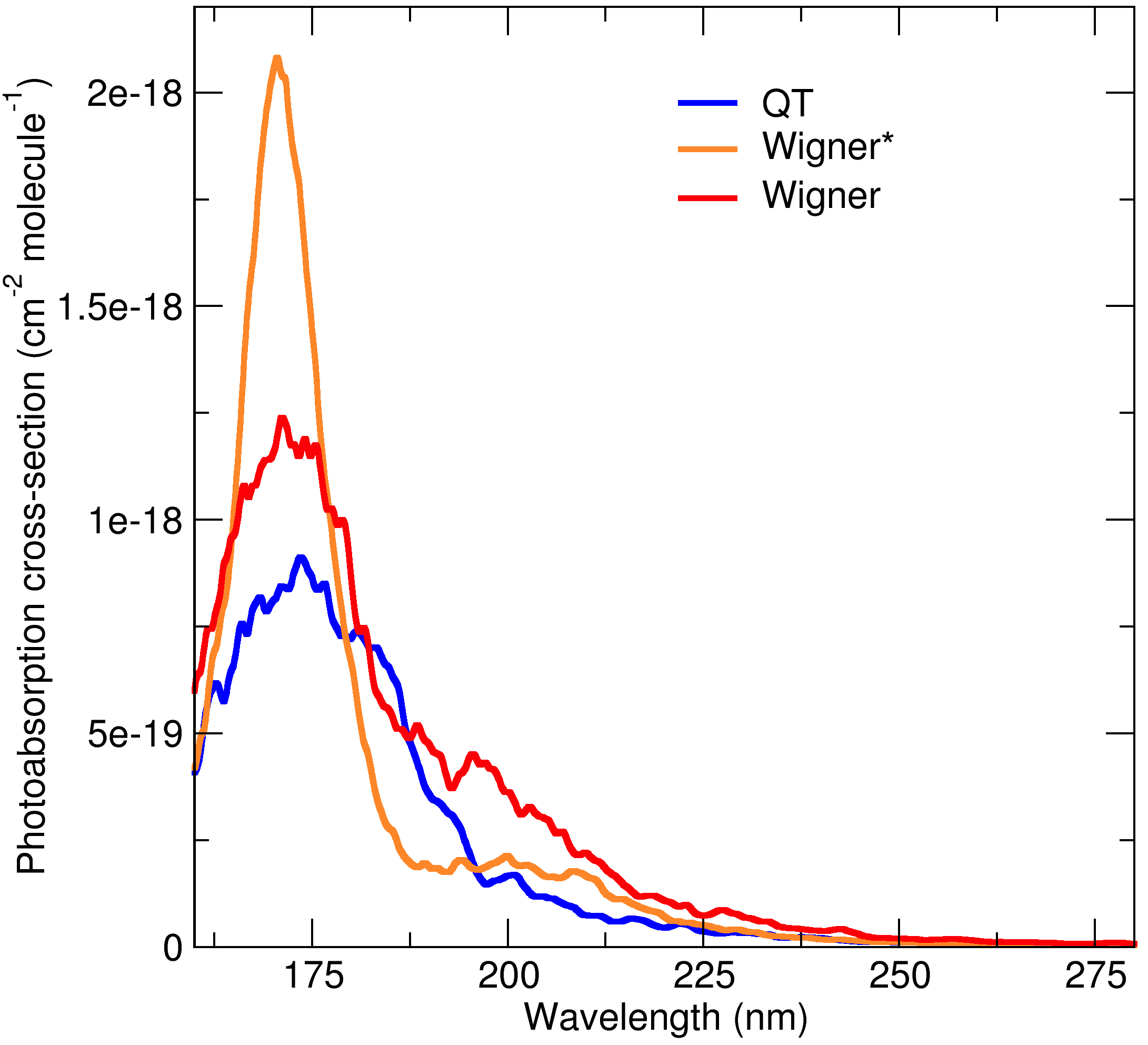}
\caption{Calculated photoabsorption cross-sections for MHP. These cross-sections are the same as those presented in the main text, but the range at short wavelength was extended to better visualize the narrower band centered at 170~nm obtained with geometries sampled from Wigner*.}
\label{fig:spectrafull}
\end{figure}

\begin{figure}[!ht]
\centering
\includegraphics[width=0.9\textwidth]{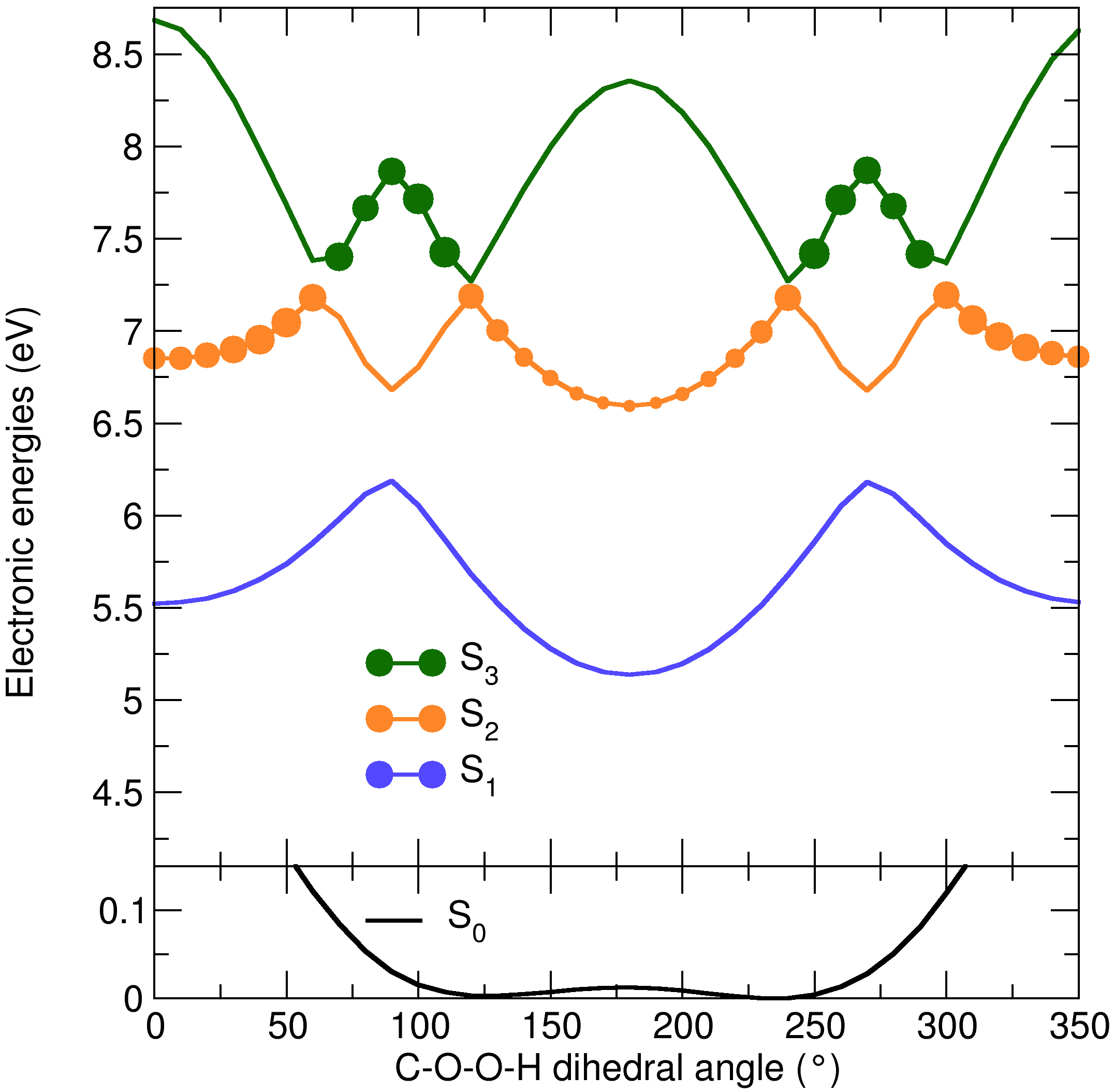}
\caption{Rigid scan along the \ce{C-O-O-H} dihedral angle of MHP. The reference geometry was obtained at the MP2/aug-cc-pVDZ level of theory. XMS-CASPT2(8/6)/def2-SVPD was used for the electronic energies and oscillator strengths. Filled circles are used to symbolize oscillator strength between S$_0$ and the (adiabatic) electronic state considered. The size of these circles is directly proportional to the size of the oscillator strength -- when the oscillator strength is negligible (for example, S$_0\rightarrow$S$_1$), only the curve for the electronic energy is visible. For sizeable oscillator strengths (for example, S$_0\rightarrow$S$_2$ or S$_0\rightarrow$S$_3$ depending on the region of the scan), the circles are clearly visible and highlight the exchange of electronic character between the adiabatic electronic states. More specifically, even a small variation of the \ce{C-O-O-H} dihedral angle around one of the two S$_0$ minima leads to an important change in the electronic energy of the electronic-state character leading to a large oscillator strength. It is important to note that the normal mode corresponding to the \ce{C-O-O-H} torsion is removed from the Wigner*, which explains the narrow high-intensity band in the Wigner* photoabsorption cross-section at around 170~nm ($\sim$ 7.3 eV) in comparison to the cross-sections obtained with the other sampling strategies (see Fig.~\ref{fig:spectrafull}).}
\label{fig:cooh-scan}
\end{figure}

\begin{figure}[!ht]
\centering
\includegraphics[width=0.9\textwidth]{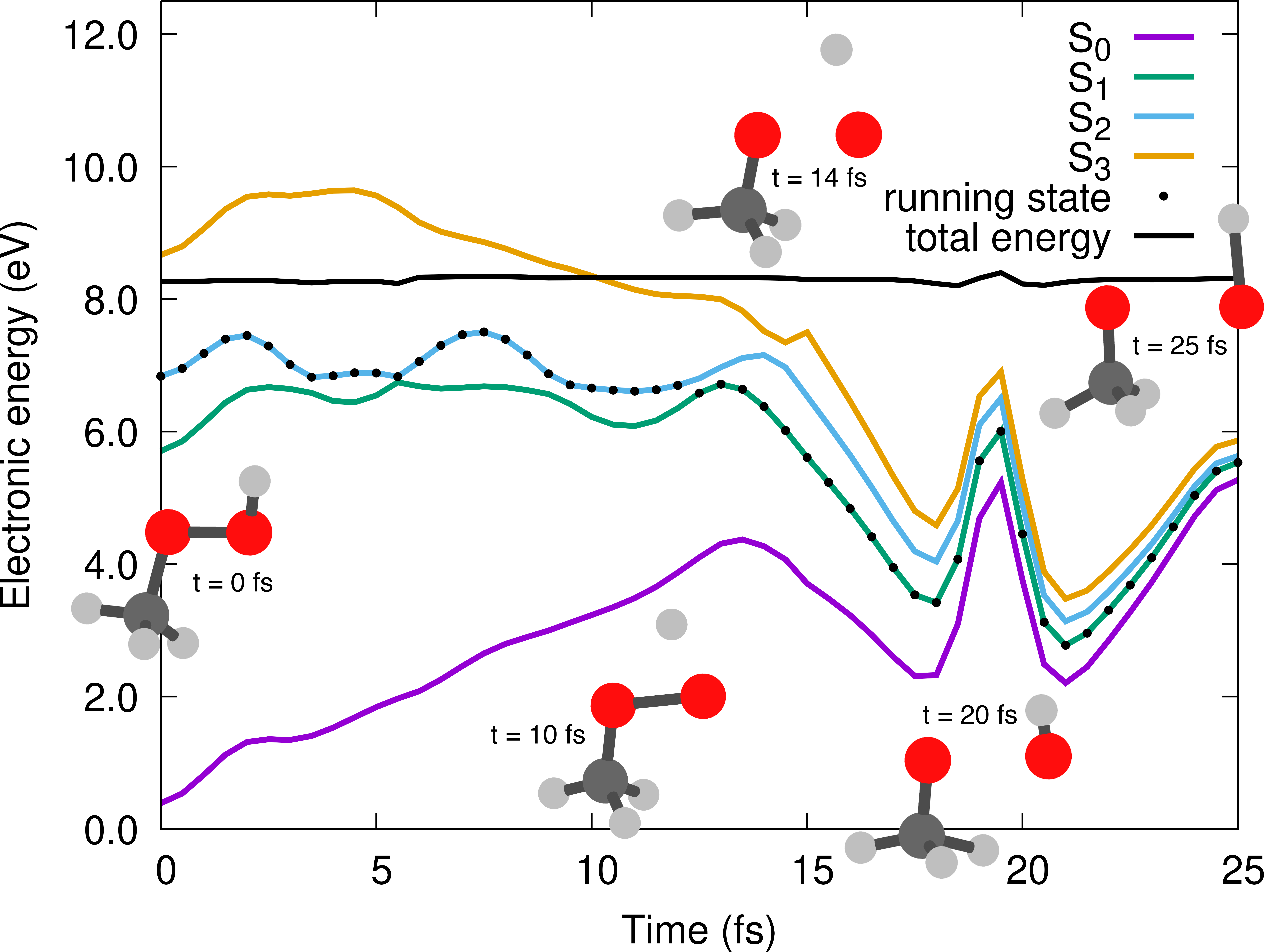}
\caption{Electronic energies along an exemplary TSH trajectory depicting the nonadiabatic process responsible for the low kinetic-energy tail in the translational kinetic energy map for OH photodissociation in the excitation window around 193~nm. MHP is initially excited into S$_2$ of the $n'\sigma^\ast$(\ce{O-H}) character and rapidly suffers an H photodissociation ($t=10$\,fs). Before the H dissociation is complete though, the molecule switches into S$_1$ of the $n'\sigma^\ast$(\ce{O-O)}) character, which triggers the \ce{O-O} bond cleavage ($t=14$ fs). The O and H then recombine ($t=20$\,fs) to form a highly vibrationally excited OH fragment, as indicated by the strong variation in electronic energy when the OH reforms between 15--20\,fs and the large amplitude oscillations of the \ce{O-H} bond in the molecular snapshots at $t=20$\,fs and $t=25$\,fs.}
\label{fig:example-traj}
\end{figure}

\begin{figure}[!ht]
\centering
\includegraphics[width=0.8\textwidth]{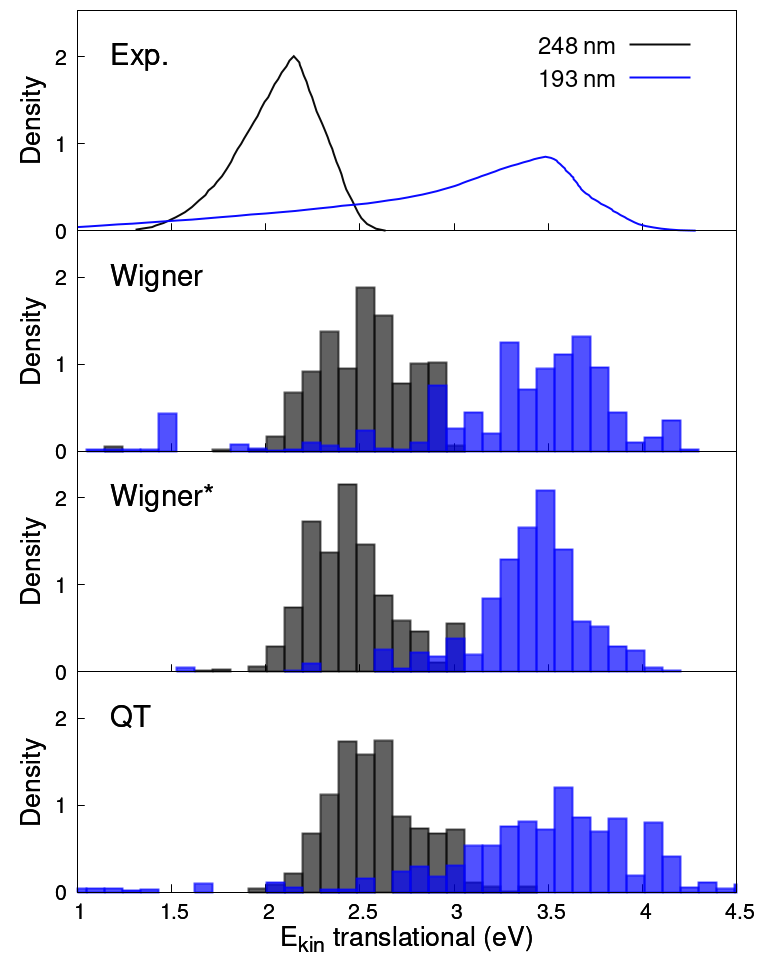}
\caption{Translational kinetic energy maps for OH photodissociation. Experimental data sets for an excitation at 248~nm (black) and 193~nm (blue) (Ref.~\citenum{thelen1993photofragmentation}) are compared to the theoretical results obtained from TSH simulations initiated from a Wigner, Wigner* or QT sampling. The initial conditions were chosen using the uniform selection, as in Fig.~5 in the main text, but the final contribution of each TSH trajectory was weighted by the initial oscillator strength (see main text for details).}
\label{fig:ekin-map-reweighted}
\end{figure} 

\clearpage
%\bibliography{peroxide}
\providecommand{\latin}[1]{#1}
\makeatletter
\providecommand{\doi}
  {\begingroup\let\do\@makeother\dospecials
  \catcode`\{=1 \catcode`\}=2 \doi@aux}
\providecommand{\doi@aux}[1]{\endgroup\texttt{#1}}
\makeatother
\providecommand*\mcitethebibliography{\thebibliography}
\csname @ifundefined\endcsname{endmcitethebibliography}
  {\let\endmcitethebibliography\endthebibliography}{}